\documentclass[aps,pre,twocolumn,floatfix,showpacs,superscriptaddress,longbibliography,nofootinbib]{revtex4-2}

\usepackage{amssymb}
\usepackage{amsmath}
\usepackage{bm}
\usepackage{graphicx}
\usepackage{hyperref}
\usepackage{xcolor}
\usepackage{soul}

\DeclareMathOperator{\Log}{Log}
\DeclareMathOperator{\sgn}{sgn}
\newcommand{\Q}{\mathbf{Q}}
\newcommand{\n}{\mathbf{\hat{n}}}
\newcommand{\zb}{\bar{z}}
\usepackage[utf8]{inputenc}

\begin{document}

\title{Structure and Dynamics of Defects in Two-Dimensional Nematic Liquid Crystals}

\author{Cody D. Schimming}
\email[]{cody.schimming@umb.edu}
\affiliation{Department of Physics, University of Massachusetts Boston, Boston, Massachusetts, 02125, USA}
\affiliation{Department of Physics and Astronomy, Johns Hopkins University, Baltimore, Maryland, 21211, USA}

\author{Daniel A. Beller}
\affiliation{Department of Physics and Astronomy, Johns Hopkins University, Baltimore, Maryland, 21211, USA}

\begin{abstract}
Topological defects in nematic liquid crystals have recently received renewed interest due to their important role in non-equilibrium active materials and developing biological systems.
While many studies have focused on simulating defect behaviors in these non-equilibrium situations, there has also been recent progress in our analytical understanding of defect morphology and behavior in a number of complicated regimes, such as intricate boundary conditions, anisotropic elasticity, or the presence of flows.
Here, we review the analytical theory of disclination structure and dynamics in two-dimensional nematic liquid crystals, highlighting recent methods and results.
\end{abstract}

\maketitle

\section{Introduction}
Topological defects are beautiful, fascinating textures in physical systems with broken symmetry.
Defects in liquid crystals are certainly no exception.
Indeed the nematic phase got its name from Georges Friedel because of the ``threadlike'' disclination lines that permeate its samples \cite{deGennes75}.
However, topological defects in liquid crystalline phases are not only visual marvels, but also important excitations that allow us to test theories in mathematical physics, statistical mechanics, and materials science.
They are also important in a variety of technological and biological applications, and have received recent renewed interest due to their importance in active materials and organism morphogenesis \cite{doo18,Maroudas-Sacks21,Balasubramaniam22}.

Even though topological defects remain one of the most commonly studied textures in liquid crystals, advancement in our analytical understanding of their structure and dynamics has been slow due to the complexity of their geometry, topology, energetics, and hydrodynamics.
A much more common approach to studying defect structures and dynamics these days comes from computational studies since the governing equations are typically highly nonlinear and nonlocal.
There have been myriad recent computational studies focusing on both equilibrium and non-equilibrium properties of defects in liquid crystals, many of which form the basis of a large body of literature on active nematics in particular.
In this review, we instead focus on the recent analytical methods and results regarding the structure and dynamics of defects in nematic liquid crystals.

\begin{figure*}
\centering
    \includegraphics[width = \textwidth]{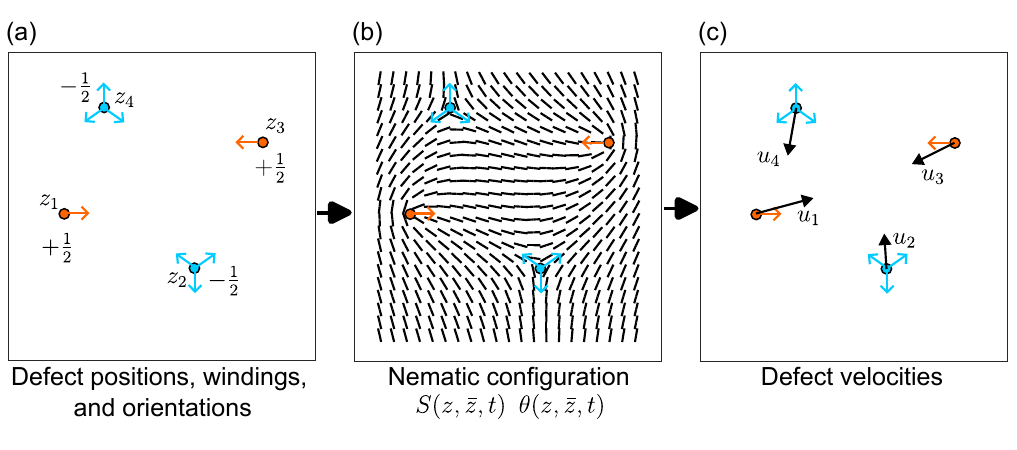}
    \caption{Visual depiction of the organizational principle of this review and our understanding of defect dynamics.
    (a) Defect positions, windings, orientations, and perhaps boundary conditions and trajectory histories are given.
    (b) The nematic configuration, i.e. the nematic order parameter $q(z,\zb,t) = Se^{i2\theta}$, is computed from the given defect parameters in (a).
    We review methods for this computation in Sec.~\ref{sec:DefectStructure}.
    (c) From the nematic configuration, defect velocities are computed.
    Methods for deriving the velocity of defects are reviewed in Sec.~\ref{sec:DefectDynamics}.
    Thus, following this work-flow, we may compute defect dynamics from given defect parameters, treating the systems as interacting particles.}
    \label{fig:DefectSchematic}
\end{figure*}

The question we seek to answer is: given the location and orientation of a number of defects in a system, what is the function that describes the nematic configuration everywhere in our domain and what is the velocity of each defect?
It turns out that answers to the first questions are essential to answering the second.
That is, one must know the nematic configuration to understand and predict the dynamics of defects.
This idea is summarized visually in Fig.~\ref{fig:DefectSchematic}.
The Holy Grail of these pursuits is to write a so-called ``particle-field model'' of defect dynamics within a nematic phase, so that we may forget about the underlying order and treat the defects as interacting particles, similarly to how we often treat interacting electrostatic charges.

Since this is such a classical topic, one may expect that most of the questions regarding defect structure and dynamics have been answered, and are now only textbook problems.
However, perhaps surprisingly, there have been many new advancements in analytical methods and results focusing on defect orientation, defect structure with differing elastic constants, defect structure in complicated domains, and new approximations to better understand defect dynamics.
Our primary goal here is to review these developments, while putting them in context of the classical results and providing a common language to bridge some of the differences in notation and style between different works.
While we will compare some analytical results to numerical results when applicable (and accessible), we will be more focused here on analytical advances and results.

With the large accessibility to computation nowadays, one may ask, why should we care so much about analytical results?
In our view, the benefits are twofold.
First, even though computation allows us to explore large systems with complex dynamics, the interpretability of such calculations is not always so clear cut.
Without the fundamental grounding that analytical results give us, relatively simple behaviors may stump us and lead to misinterpretations of data and results.
Additionally, analytical results are often much more easily manipulated allowing for wider applicability, and the limits under which they are applicable are also much easier to understand.
Analytical results in nematics may even be useful for springboarding studies of other broken symmetry phases in fields as far-reaching as cosmology \cite{chuang91,bowick94}. 
Second, analytical results do not necessarily compete with computations, but can open new avenues of computational research.
Having a ``particle theory'' of defect dynamics would allow us to simulate many more defects than could be done with continuum theory given the same computational power.
Additionally, incorporating analytical results in the computation itself (or the visualization of the computation) may speed up the process immensely.

It is often the case that a theoretical discussion of \textit{topological} defects revolves around \textit{topology}.
Here, however, we will avoid such discussions insofar as is required to understand the definition of a topological defect.
There are several excellent classical reviews (see e.g.  \cite{Mermin79,kleman89,kleman08,alexander12}) on the use of topology in materials physics and nematics.
While there are still unanswered questions (as well as plenty of recent progress on these questions) surrounding the topological aspects of defects in liquid crystalline phases, we will leave a proper review of these for the future.
Further, we limit ourselves in this review to two dimensions where the topology is fairly straightforward.
There are certainly enough recent results for two-dimensional defect dynamics to fill a whole review, so we leave the issue of three-dimensional defect dynamics to a future review (though we will briefly mention some of the outstanding problems related to three dimensions, and how they differ from two dimensions, at the end of this article).

The plan for this review is as follows: In Sec.~\ref{sec:TopDefects} we will review the definition of a topological defect and give a very brief discussion of what makes them \textit{topological}.
We will also introduce the nematic director and order parameter, as well as give some mathematical background for the rest of the paper.
In Sec.~\ref{sec:DefectStructure} we will discuss analytical solutions for the nematic configuration (that is, the director and order parameter) when defects are present as well as highlight some of the methods used to obtain the results.
In Sec.~\ref{sec:DefectDynamics} we will review several techniques for taking the solutions of Sec.~\ref{sec:DefectStructure} and translating them into equations for the defect velocity.
Throughout both Secs.~\ref{sec:DefectStructure} and \ref{sec:DefectDynamics} we will rely heavily on complex-valued coordinates which not only simplify the expressions in most cases, but also heavily lend themselves to the mathematical techniques that are used to derive the expressions.
To this end, we will attempt throughout the review to stick to complex-valued coordinates and notation, and will explicitly state when real-valued coordinates are used.
Finally, in Sec.~\ref{sec:Conclusion} we conclude with a summary as well as future challenges surrounding our analytical understanding of nematic defects.

\section{Topological Defects in Nematic Phases} \label{sec:TopDefects}

\subsection{Classification of Topological Defects}
Topological defects result from broken continuous symmetries in a physical system.
In the nematic phase, orientational symmetry is broken and the local orientation is often described with a unit vector called the director $\n$.
Importantly, nematic phases differ from polar phases (such as a classical ferromagnet) in that the physical state of the system is invariant under rotations by $\pi$, that is, $\n\to-\n$ does not change a given configuration.
By configuration, we mean an assignment of a director for every point in space, $\n(\mathbf{r})$.

Since we are working in two dimensions, it will be prudent to use complex-valued coordinates.
In many cases, this allows for more compact and elegant notation.
Additionally, several aspects of complex analysis will aid in our development of analytical methods.
To this end, we introduce here the complex coordinate system with (independent) complex coordinates $z = x +iy$ and $\zb = x - iy$.
Partial derivatives with respect to these coordinates may be translated into derivatives with respect to $x$ and $y$ from the Wirtinger derivatives:
\begin{align}
    \partial_z &= \frac{1}{2}\left(\frac{\partial}{\partial x} - i\frac{\partial}{\partial y}\right) \\
    \partial_{\zb} &= \frac{1}{2}\left(\frac{\partial}{\partial x} + i \frac{\partial}{\partial y}\right). 
\end{align}
Complex-valued functions, which represent two-dimensional vectors (likewise, real-valued functions represent scalars), admit the following differential operations:
\begin{align}
    \nabla\cdot\mathbf{f} &\equiv 2\Re\left[\partial_z f\right] \\
    \nabla \times\mathbf{f} &\equiv 2\Im\left[\partial_z f\right] \\
    \nabla^2\mathbf{f} &\equiv 4\partial_z\partial_{\zb}f
\end{align}
where $\Re[\cdot]$ ($\Im[\cdot]$) represents the real (imaginary) part of a complex number.
Finally, when a complex-valued (or real-valued) function is integrated over a domain $\Gamma$, it is common to use the following notation
\begin{equation}
    \int_{\Gamma} f(x,y) \,dx dy \equiv \frac{i}{2}\int_{\Gamma} f(z,\zb) \, dz d\zb
\end{equation}
where the factor of $i/2$ is the Jacobian of the transformation from $\{x,y\}$ coordinates to $\{z,\zb\}$ coordinates.

Though there are mathematical benefits to using complex-valued coordinates, one of the drawbacks is that traditional definitions and results must be expressed in a potentially unfamiliar language.
Readers may compare more traditional expressions using vector analysis with their complex-valued alternatives by expressing the real part as the $x$ component while the imaginary part as the $y$ component.
The director in complex-valued coordinates is given by $n(z,\zb) = e^{i\theta(z,\zb)} = \cos\theta + i\sin\theta$ where $\theta$ is the angle of the director with respect to the real axis ($x$ axis).
As with the real-valued director defined above, $n$ and $-n$ represent the same physical state.

The simplest energy of a configuration that one might write down is the Oseen-Frank elastic energy \cite{frank58}:
\begin{align} \label{eqn:FrankEnergy}
    F_{OF} &=\int_\Gamma \left[\frac{K_{11}}{2}(\nabla\cdot\n)^2 + \frac{K_{33}}{2}(\nabla\times\n)^2 \right]\,dxdy \\
    &=\frac{i}{2}\int_{\Gamma}\left[ 2K_{11}\Re[\partial_zn]^2 + 2 K_{33}\Im[\partial_z n]^2\right] \, dzd\zb
\end{align}
where $\Gamma$ is the domain and the $K$'s are elastic constants corresponding to splay and bend \cite{selinger19}.
The minimum energy state of a pure nematic phase is a uniform configuration, $n(z,\zb) = n_0$.
Small, continuous perturbations of a uniform configuration cost energy, and hence typically decay to a uniform state unless they are required by boundary conditions.

\begin{figure}
\centering
    \includegraphics[width = \columnwidth]{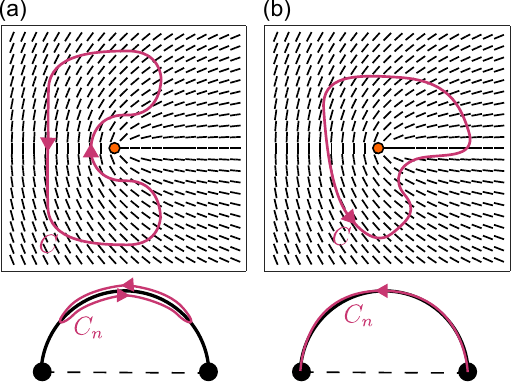}
    \caption{Visual example of the topological description of a defect.
    (a) The measuring curve $C$ does not enclose the defect, so the corresponding curve $C_n$ in configuration space may be continuously shrunk to a point.
    (b) If $C$ encloses the defect, the corresponding curve $C_n$ in configuration space winds around the entire space without doubling back, so we cannot continuously deform it to a point.
    This is the topological definition of a defect.}
    \label{fig:DefectTopology}
\end{figure}

Loosely speaking, a topological defect in the nematic phase is an excitation of the ordered (ground) state which cannot be removed by any continuous transformation (holding the location of the defect fixed).
How do we know whether a configuration can or cannot be removed by a continuous transformation?
To answer this question, we must resort to the field of algebraic topology, and, in particular, we must ask a question about loops in configuration space.
By configuration space, we mean the space in which the director $n$ lives (sometimes called the order parameter space or ground state manifold).
In two dimensions, the configuration space can be visualized as a semi-circle with endpoints identified (since $n$ and $-n$ represent the same state), usually called the ``real projective line,'' $\mathbb{RP}^1$.
If we take a loop $C$ in our domain, the director along that loop gives us a corresponding loop $C_n$ in the configuration space.
If $C_n$ can be continuously shrunk to a point, then the texture surrounded by $C$ does not contain a defect; on the other hand, if $C_n$ cannot be continuously shrunk to a point (because it maps to the whole of $C_n$ and does not double back on itself), then a defect lies somewhere within the region surrounded by $C$.
In Fig.~\ref{fig:DefectTopology}, we provide a visual example of this process.
We note that the topology of defects in three dimensions is quite a bit more complicated, and point readers to~\cite{Mermin79,stark01,Kleman06,alexander12,Wu22} for more detailed discussions.

In two-dimensional nematics the measuring curve $C$ may correspond to a configurational space curve $C_n$ that wraps around configurational space multiple times.
The number of times this curve wraps around the space measures the winding number of the defect (or collection of defects) enclosed in the curve.
Writing $n = e^{i\theta}$, the winding number may be directly computed by
\begin{equation} \label{eqn:2DWindingNumber}
    m = \frac{1}{2\pi}\oint_C \nabla\theta\cdot d\bm{\ell} = \frac{1}{2\pi}\oint_C \partial_z\theta\,dz.
\end{equation}
Since $n\equiv-n$, $m = k/2$ for $k$ an integer.
In two dimensions, defects may be distinguished by their winding number.
By using measuring curves that surround multiple defects we find that composite winding numbers are just the sum of individual winding numbers, indicating that opposite signed winding defects may annihilate one another.

\subsection{Nematic Order Parameter and Free Energy}

Given a measuring curve from the previous section, if we now deform the curve so that it gets smaller and smaller, eventually the curve will pass over a point which changes the measured winding number.
This point is the location of the defect.
In our description of nematics above in terms of the director field $n(z,\zb)$, this point represents a singularity: the energy density, the integrand of Eq.~\eqref{eqn:FrankEnergy}, diverges, and $n(z_0)$ (where $z_0$ is the location of the defect) is undefined.
One method of rectifying this issue is to ``cut out'' or exclude a region of some small radius $\varepsilon$ around the location of the defect.
The energy of the defect configuration can then be computed (it will be a function of $\varepsilon$) and there are no longer any undefined points in the included domain.

While the cutout method allows energies of configurations to be computed, it is impractical for many studies of defect dynamics, both analytical and numerical.
Indeed, in most of the methods for approximating defect dynamics that we review below, some knowledge of what the configuration looks like near the location of the defect (i.e. distances smaller than $\varepsilon$) is required.
How can we write a description of the configuration that is continuous even at defects?

Introduce the scalar order parameter $s(z,\zb)$.
The scalar order parameter represents the degree of alignment of the microscopic nematogens of the liquid crystal, and hence is a statistical quantity.
Typically, $s = 1$ represents a perfectly ordered state, while $s=0$ represents a state with a completely random distribution of nematogen orientations.
Therefore, in a typical nematic configuration with defects, we expect $s\approx 1$ away from defects ($|\mathbf{r}-\mathbf{r}_0| >\varepsilon$), but $s \to 0$ as $|\mathbf{r}-\mathbf{r}_0|\to0$ so that the singularity may be regularized.
We note that the situation in three-dimensional nematics is more complicated, since the core may become biaxial \cite{lyuksyutov78,schopohl87}.

We can organize the information from $n$ and $s$ together into a single mathematical object, which is called the nematic order parameter $q(z,\zb) = s(z,\zb) e^{i2\theta(z,\zb)}$.
Thus, the magnitude of $q$ gives information about the local degree of ordering, while the argument of $q$ (the complex polar angle) gives information about the orientation.
An additional benefit of describing the nematic state using the order parameter is that the factor of two in the exponential part of $q$ ensures that nematic symmetry holds.
That is, if $\theta \to \theta + \pi$ (or $n \to -n$), $q$ remains the same.
We note that the complex-valued order parameter $q$ is related to the tensor-valued order parameter $\Q = 2s(n_i n_j - \tfrac{1}{2} \delta_{ij})$ that is often used for real-valued coordinate systems \cite{deGennes75} by $q = Q_{11} + i Q_{12}$ so that there is a simple, direct mapping between both descriptions.

There are already several excellent reviews using real-valued coordinates on $\Q$-tensor theory \cite{gramsbergen86,James08,mottram14}.
For our purposes, the important aspects are the free energy and dynamical equations, which we will write in complex-valued coordinates here for consistency.
The free energy may be written generally as
\begin{multline} \label{eqn:QFreeEnergy}
    \mathcal{F}_q = \frac{i}{2}\int_\Gamma \left[f_b(q) + f_e(q,\partial_z q) + f_H(q,H)\right]\,dz d\zb \\
    + \oint_{\partial\Gamma} f_S(q)\, dz
\end{multline}
where $f_b$ is a bulk free energy density determining the ground state phase (i.e. the equilibrium values of $s$), $f_e$ is an elastic energy density that penalizes non-uniform configurations, $f_H$ is a free energy density associated with the application of an external field $H$, and $f_S$ is a surface (or ``anchoring'') free energy per unit length that only applies to the boundary of the domain.
$f_b$ is often taken as the double-well (often called ``Landau de-Gennes'' when the tensor order parameter is used) potential:
\begin{equation} \label{eqn:Landau-deGennes}
    f_b = \frac{A}{2}|q|^2 + \frac{C}{4}|q|^4
\end{equation}
where the equilibrium phase is determined by the sign of $A$ ($A < 0$ for nematic and $A > 0$ for isotropic) and $C$ is a phenomenological energy scale.
The equilibrium value of $s$ is given by $s_0 = \sqrt{-A/C}$ when the system is in the nematic phase.
The elastic energy has many possible terms, depending on the elastic constants in Frank-Oseen theory [Eq.~\eqref{eqn:FrankEnergy}] that need to be modeled \cite{longa87,selinger19}.
In the common ``single elastic constant'' approximation, where the Oseen-Frank free energy density can be written $\tfrac{1}{2} K (\partial_i n_j) (\partial_i n_j) $ with $i$ and $j$ both summed over $x$ and $y$, the corresponding expression in complex coordinates is simply
\begin{equation} \label{eqn:QOneConstantElastic}
    f_e = 2K|\partial_z q|^2.
\end{equation}
We will briefly consider other choices of $f_e$ in later sections, but most of this review will assume a one constant approximation corresponding to Eq.~\eqref{eqn:QOneConstantElastic}.
We will not worry about the terms $f_H$ and $f_S$ in this review, and we note that there is a great deal of literature already exploring these terms \cite{Meyer69,sen86,gramsbergen86,Nobili92,Fournier05,Willman07,James08,mottram14}.

Finally, the dynamics of $q$ will be important for the dynamics of defects.
In the presence of a fluid flow $v = v_x + iv_y$, we will assume $q$ evolves according to the Beris-Edwards equations \cite{beris94}
\begin{equation} \label{eqn:BerisEdwards}
    \frac{\partial q}{\partial t} + \left(v\partial_z+\bar{v}\partial_{\zb}\right)q = \Omega -\frac{1}{\gamma}\frac{\delta \mathcal{F}}{\delta \bar{q}}
\end{equation}
where $\gamma$ is the rotational viscosity and
\begin{multline} \label{eqn:STensor}
    \Omega = \lambda\partial_{\zb} v + q(\partial_zv - \partial_{\zb} \bar{v}) -4\lambda q(q\partial_zv + \bar{q}\partial_{\zb}\bar{v}) 
\end{multline}
is a generalized advection for anisotropic fluids.
The parameter $\lambda$ encodes the tendency for the anisotropic particles to align with (or tumble in) shear flows.
In the absence of flow, Eq.~\eqref{eqn:BerisEdwards} reduces to $\gamma\partial q/\partial t = -\delta\mathcal{F}_{q} /\delta\bar{q}$, i.e. gradient flow of the free energy.
The steady state in this case will be a local minimum of Eq.~\eqref{eqn:QFreeEnergy}.

\section{Defect Structure} \label{sec:DefectStructure}

Our goal in describing the structure of defects is to write down a function of spatial coordinates that describes the order parameter, subject to the constraint that Eq.~\eqref{eqn:2DWindingNumber} holds for appropriate choices of measuring curves, given the location of defects in the system.
That is, at a snapshot in time, what is the order parameter?
Of course, if the system has boundaries, the function must also satisfy the boundary conditions.
However, these conditions are not actually enough to uniquely define the configuration.
There are infinite families of functions that one could write down that satisfy Eq.~\eqref{eqn:2DWindingNumber}.
To further constrain what we are looking for, we will also require that the function be a local minimum of the free energy, subject to the constraints of defect locations and boundary conditions.
In the absence of flow, this condition guarantees that the configuration will be a stationary state (that is, $\partial_t q = 0$) under the mentioned constraints. We will refer to such a state as a \textit{static configuration} because the defects are pinned in place. 
However, if defects are able to move, their rearrangements become part of the relaxation dynamics.  
A common approximation (which we will explore in more depth in Sec.~\ref{sec:DefectDynamics}) is that the configuration relaxes much faster than the defect moves, so that the static configuration may be used to describe the director and order parameter instantaneously based on the defect's current location.
Our approach in this section is to compute static configurations with this approximation in mind.
Of course, there can also be scenarios (e.g. when a boundary condition requires a number of topological defects) where defects are truly static, and these configurations should be described by the methods and results of this section.

There is one more issue we will touch on: that of far-field and near-field defect structure.
As mentioned above, in the far-field (far away from the location of any defect), we may assume $s\approx s_0$, a constant set by the bulk free energy density.
For convenience,  we will rescale $s$ and $q$ by $s_0$ and define $S = s/s_0$ so that far from the core of defects $|q| = S = 1$.
In this limit, it is sufficient to find the director, and hence we may look for local minima of Eq.~\eqref{eqn:FrankEnergy} (the Oseen-Frank free energy).
In the near-field limit, however, we must know both the director and scalar order $S$, since $S\to0$ at the defect core.
Having at least an approximation of both of these limits will be required for methods that give approximate velocities of defects in the next section.

\subsection{Single Elastic Constant Far-Field Multi-Defect Configuration}
We begin by exploring the most common theoretical situation: $N$ defects located at $\{z_i\}$ with ``free'' (or infinite) boundary conditions.
Starting with the far-field (director) configuration, we will also assume that the elastic constants in Eq.~\eqref{eqn:FrankEnergy} are all equal (often referred to as the ``one constant approximation'').
The Oseen-Frank energy can then be written
\begin{equation} \label{eqn:1Const2DFrank}
    \mathcal{F}_{OF} = \frac{i}{2}\int_\Gamma 2K|\partial_zn|^2\,dz d\zb = \frac{i}{2}\int_\Gamma 2K|\partial_z\theta|^2 \, dz d\zb
\end{equation}
where the second equality is written in terms of the angle $\theta$ by substituting $n = e^{i\theta}$.
To be a local minimum of the free energy, the angle $\theta(z,\zb)$ must satisfy the Euler-Lagrange equation
\begin{equation} \label{eqn:2DElEquation}
    \partial_{\zb} \partial_z\theta = 0
\end{equation}
or, in real-valued coordinates $\nabla^2\theta = 0$.
That is, $\theta$ must be a harmonic function.

The Euler-Lagrange equation for $\theta$ has a couple nice properties.
First, the equation is linear, so we only need to consider a solution for a single defect.
The full solution for multiple defects will just be the sum of all the single defect solutions.
Second, for complex-valued functions, any analytic function (that is, a differentiable function that does not depend on $\zb$) will have real and imaginary parts that are harmonic.
So, we may re-frame our search for an analytic function $\Phi(z)$ such that $\theta(z,\zb) = \Im\left[\Phi\right]$ where $\Im\left[\cdot\right]$ denotes the imaginary part of a complex valued function.
Note that we could also define $\theta(z,\zb) = \Re\left[\Phi\right]$ (where $\Re\left[\cdot\right]$ is the real part), and the literature is somewhat split between these two conventions.
We choose the imaginary part for physical consistency when comparisons to electrostatics are made, and for notational convenience when considering solutions to Eq.~\eqref{eqn:2DElEquation}.
The harmonic conjugate of $\theta$, i.e. $\phi = \Re\left[\Phi\right]$, may be interpreted as the electrostatic potential of the equivalent electrostatic configuration \cite{Vitelli06,Hill21}.
A useful relationship between $\theta$ and $\phi$ are the Cauchy-Riemann equations
\begin{align}
    \frac{\partial\phi}{\partial x} &= \frac{\partial \theta}{\partial y} \nonumber \\
    \frac{\partial\phi}{\partial y} &= -\frac{\partial \theta}{\partial x} \label{eqn:CREquations}
\end{align}
which hold if $\Phi(z)$ is an analytic function.
We note that defects (or electrostatic charges) will appear as singularities in $\Phi(z)$. 

To satisfy the winding number condition, Eq.~\eqref{eqn:2DWindingNumber}, $\theta$ must be multivalued.
For a single defect, the function $\Phi(z) = m \Log\left[z - z_0\right] = m\log(|z-z_0|)+im\arg(z-z_0)$ is analytic and its imaginary part is multivalued, as the value of $\arg(z-z_0)$ is only unique modulo $2\pi$. Furthermore, the imaginary part $\theta$ has a singularity at $z = z_0$, so that 
\begin{equation} \label{eqn:ComplexSingleDefectDirector}
\theta(z,\zb) = m\arg\left(z - z_0\right) + \theta_0    
\end{equation}
or, in real-valued coordinates
\begin{equation} \label{eqn:RealSingleDefectDirector}
    \theta(x,y) = m\arctan\left(\frac{y - y_0}{x - x_0}\right) + \theta_0,
\end{equation}
where $\theta_0$ is a constant representing a global phase that does not change the energy of the configuration.
Equation \eqref{eqn:ComplexSingleDefectDirector} was used to generate the director field shown in Fig.~\ref{fig:DefectTopology} with $m = 1/2$ and $\theta_0 = 0$.
Substituting Eq.~\eqref{eqn:ComplexSingleDefectDirector} into Eq.~\eqref{eqn:1Const2DFrank}, the energy of a single defect is 
\begin{equation}
\mathcal{F}_{d}=\int_0^{2\pi}\int_\varepsilon^{R}\frac{K}{2}\frac{m^2}{r^2}\,rdr d\varphi + F_c =\pi K m^2\log\left(\frac{R}{\varepsilon}\right) + F_c
\end{equation}
where $R$ is a length scale characterizing the system size, $\varepsilon$ is the defect core size, and $F_c$ is the energy of the core region ($r < \varepsilon$) which cannot be computed with only the director.
Note that the energy diverges if we take $R \to \infty$ or $\varepsilon\to 0$.

For an arbitrary number of defects with charges $m_i$ and locations $z_i$, the director angle is given by
\begin{equation} \label{eqn:ManyDefectDirector}
    \theta(z,\zb) = \sum_i m_i \arg\left(z - z_i\right) + \theta_0.
\end{equation}
To compute the Oseen-Frank energy of the multidefect configuration, it is useful to first rewrite the energy using Green's identity (i.e. the divergence theorem) \cite{Copar24}:
\begin{multline} \label{eqn:ContourFreeEnergy}
    \mathcal{F}_{OF} = \frac{K}{2}\int_\Gamma \nabla\cdot\left(\theta\nabla\theta\right)\,dz d\zb \\ = \frac{K}{2}\oint_{\partial\Gamma} \theta \left(\nabla\theta\cdot\bm{\hat{\nu}}\right)\, d\ell = \frac{K}{2}\oint_{\partial \Gamma} \theta\left(\nabla \phi\cdot\mathbf{\hat{t}}\right) \, d\ell \\= -\frac{K}{2}\oint_{\partial\Gamma} \theta \,d\phi = \frac{K}{2}\oint_{\partial\Gamma}\phi\, d\theta
\end{multline}
where $\oint_{\partial\Gamma}$ indicates a loop integral around the ``boundary'' of the domain, including around branch cuts and defect cores. 
The first equality of Eq.~\eqref{eqn:ContourFreeEnergy} follows from the fact that $\theta$ is harmonic, and the third and fifth equalities follow from the Cauchy-Riemann equations, Eq.~\eqref{eqn:CREquations}.

\begin{figure}
\centering
    \includegraphics[width = \columnwidth]{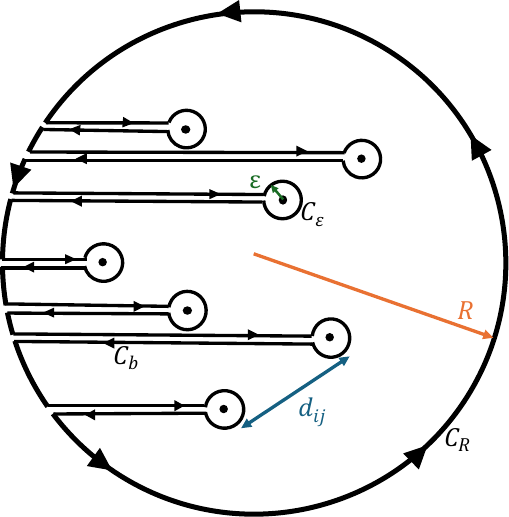}
    \caption{An example keyhole integration contour depicting defect locations and $\partial\Gamma$.
    $\partial\Gamma$ is broken up into pieces $C_R$, $C_{b_i}$, and $C_{\varepsilon_i}$. 
    $C_R$ is a circle of radius $R$ that encompasses all defects in the system.
    $C_{b_i}$ are pieces of the curve along branch cuts for each defect.
    $C_{\varepsilon_i}$ are small circles of radius $\varepsilon$ which encircle individual defects.}
    \label{fig:KeyholeIntegration}
\end{figure}

We may now evaluate the energy by computing a keyhole integration over a curve such as shown in Fig.~\ref{fig:KeyholeIntegration} in the following manner.
We break the contour $\partial\Gamma$ up into pieces: $C_R$, a circle of radius $R$ that encompasses all defects in the system and several smaller pieces: segments $C_{b_i}$ which go both ways from defect to defect along branch cuts, and segments $C_{\varepsilon_i}$ which form circles of radius $\varepsilon$ (the defect core radius) around each defect.
We also define $d_{ij} = |z_i -z_j|$, the distance between defects $i$ and $j$, $\theta_{ij} = \arg(z_i - z_j)$, the angle between defects $i$ and $j$, $d^* = \max\left(d_{ij}\right)$, the longest distance between defects, and $d_* = \min\left(d_{ij}\right)$, the shortest distance between defects.
Using the last equality of Eq.~\eqref{eqn:ContourFreeEnergy} and noting that
\begin{equation}
    \phi(z,\zb) = \sum_i m_i \log\left|z - z_i\right|
\end{equation}
we calculate
\begin{equation} \label{eqn:CRIntegral}
    \oint_{C_R} \phi \,d\theta = 2\pi\left(\sum_i m_i\right)^2\log\left(R\right) + O\left(\frac{d^*}{R}\right)
\end{equation}
since $\phi = \sum_i m_i\log\left(R\right) + O(d^*/R)$ and $\theta = \sum_j m_j \arg(z) + O(d^*/R)$ along this part of the contour.
The contour integrals for each piece $C_{b_i}$ are zero since there are equal pieces going in opposite directions that cancel (note that $d\arg(z)$ is an odd function of $z$ while $\log|z|$ is even). 
Finally, on the contours $C_{\varepsilon_i}$ for the $i$th defect, we have 
\begin{align}
    \phi &= m_i \log \varepsilon + \sum_{j\neq i}m_j\log d_{ij} + O\left(\frac{\varepsilon}{d_*}\right) \\
    \theta &= m_i\arg(z - z_i) + \sum_{j\neq i} m_j \theta_{ij}+ O\left(\frac{\varepsilon}{d_*}\right)
\end{align}
so that the sum over all contours $C_{\varepsilon_i}$ is (keeping in mind the circles run in the negative azimuthal direction)
\begin{multline} \label{eqn:CEpsIntegral}
    \sum_i\oint_{C_{\varepsilon_i}} \phi \, d\theta = -2\pi\sum_i\left[m_i^2 \log\varepsilon + \sum_{j\neq i}m_i m_j\log d_{ij}\right] \\+ O\left(\frac{\varepsilon}{d_*}\right).
\end{multline}
It is common to assume the separation of length scales $R \gg d^*$ and $d_* \gg \varepsilon$ so that the terms $O(d^*/R)$ and $O(\varepsilon/d_*)$ may be dropped \cite{Copar24}.
Combining Eqs.~\eqref{eqn:CRIntegral} and \eqref{eqn:CEpsIntegral}, adding and subtracting the term $2\pi\sum_i\sum_{j\neq i} m_i m_j \log \varepsilon$ and factoring gives the multi-defect energy:
\begin{multline} \label{eqn:MultiDefectFO}
    \mathcal{F}_{OF} = \pi K\left\{\left(\sum_i m_i\right)^2 \log\left(\frac{R}{\varepsilon}\right)\right. \\\left.- \sum_{i,j\neq i}m_i m_j\log\left(\frac{d_{ij}}{\varepsilon}\right)\right\}.
\end{multline}
We note that here we have assumed that every defect has the same core size $\varepsilon$, though the calculation may be generalized so that each defect has a different core size \cite{Copar24}.

Equation \eqref{eqn:MultiDefectFO} has two pieces, a defect self energy and an interaction energy.
As with the single defect, the self energy term may diverge as $R \to \infty$, but only if the total winding number is different than zero.
If the total winding number is zero, the far field director will be uniform.
The interaction energy depends on the distances between defects and the relative charges of the defects, just as in the two-dimensional electrostatic analogy.
In fact, Eq.~\eqref{eqn:MultiDefectFO} is also the energy of a system of two-dimensional point charges if one replaces the winding numbers $m_i$ with electrostatic charges $q_i$.

An important feature of the energy in Eq.~\eqref{eqn:MultiDefectFO} is that it scales with the square of the winding numbers.
This means that, given some fixed total winding, it will cost less energy to have defects of smaller winding, even if that means having more defects total.
That is, it is predicted, and seen in experiments \cite{deGennes75}, that half-integer ($m = \pm 1/2$) defects populate the system and higher winding defects split or decay into smaller winding defects (unless stabilized by boundary conditions or external fields).

\subsection{Defect orientation}

Given defect locations and charges, have we fully described the director configuration in Eq.~\eqref{eqn:ManyDefectDirector}?
The answer is no!
There is one other degree of freedom that defects possess which we have not yet mentioned: orientation (or intrinsic phase).
Equation \eqref{eqn:ComplexSingleDefectDirector}, which describes the director for a single defect, includes a global phase $\theta_0$.
This phase is also the phase of the single defect, defined as $\theta_i =\theta(z = z_i + \varepsilon)$, that is, the angle of the director at the positive $x$ position just outside the defect core.
Note, however, that the global phase in Eq.~\eqref{eqn:ManyDefectDirector} cannot be set so that each $\theta_i$ is independently determined; there are not enough degrees of freedom.

The first theoretical work that addressed orientations of defects as independent degrees of freedom was by Vromans and Giomi~\cite{vromans16} who studied the structure and interaction of two defects with arbitrary orientation using an ansatz for the director structure (a similar ansatz was used in a later study by Pearce and Kruse \cite{pearce21}).
Later, Tang and Selinger~\cite{tang17} analytically computed the director field for two defects with arbitrary orientations by minimizing the free energy subject to boundary conditions that corresponded to the defect orientations.

\begin{figure}
\centering
    \includegraphics[width = \columnwidth]{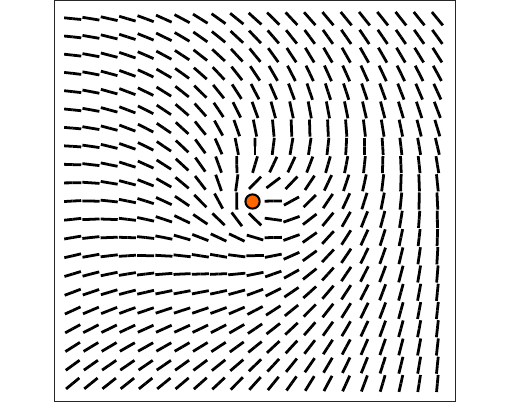}
    \caption{A defect with winding $m = +1/2$ and spiral charge $\mu = 0.7$.
    The director field acquires a radius dependent rotation due to the logarithmic addition, leading to the appearance of a spiral.}
    \label{fig:SpiralChargeDefect}
\end{figure}

The generalization to arbitrarily many defects would come from \v{C}opar and Kos~\cite{Copar24}.
The key realization is that an energy minimizing configuration (subject to the boundary conditions that give rise to each defect orientation) must be harmonic, yet we know that both $\theta$ and $\phi$ for the defect configuration are harmonic functions.
Therefore, we can introduce a generalized director configuration:
\begin{equation} \label{eqn:MultiDefectSpiralCharge}
    \theta(z,\zb) = \theta_0 + \sum_i m_i \arg\left(z - z_i\right) + \sum_i \mu_i \log\left|z - z_i\right|
\end{equation}
where the logarithm is inspired by the electrostatic potential $\phi$, leaving $\theta_0$ and $\{\mu_i\}$ as constants which will be determined by defect orientations. Since $\nabla^2\theta = 0$, Eq.~\eqref{eqn:MultiDefectSpiralCharge} is guaranteed to be a minimizer of the Oseen-Frank free energy, at least locally in the space of functions $\theta(z,\bar z)$.
For a single defect, a nonzero $\mu$ will give the director field a spiral pattern, which we show in Fig.~\ref{fig:SpiralChargeDefect} ($m = 1/2$, $\mu = 0.7$).
For this reason, the $\{\mu_i\}$ were given the name ``spiral charges'' \cite{Copar24}, which we will use going forward.
We note that, unlike the winding number $m$, the spiral charge is not quantized and can take on any real value.

The free energy corresponding to Eq.~\eqref{eqn:MultiDefectSpiralCharge} may be computed exactly the same way as above since $|\nabla\theta|^2 = |\nabla\phi|^2$.
This leads to the full free energy:
\begin{multline} \label{eqn:SpiralChargeFreeEnergy}
    \mathcal{F}_{OF} = \pi K\left\{\left[\left(\sum_i m_i\right)^2 + \left(\sum_i \mu_i\right)^2\right] \log \left(\frac{R}{\varepsilon}\right) \right. \\
    \left. - \sum_{i,j\neq i}\left(m_i m_j + \mu_i \mu_j\right)\log\left(\frac{d_{ij}}{\varepsilon}\right)\right\}
\end{multline}
so that (for this hierarchy of length scales) the contribution from winding numbers and spiral charges completely decouple.

\begin{figure}
\centering
    \includegraphics[width = \columnwidth]{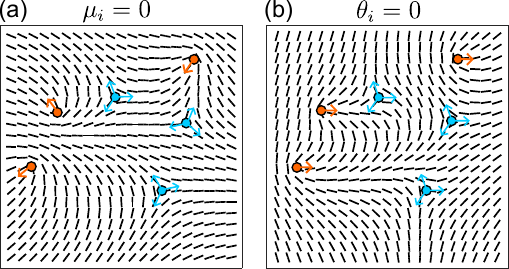}
    \caption{Comparison of two nematic configurations with defects in the same location with either (a) spiral charges $\mu_i = 0$ (and global phase $\theta_0 = 0$) or (b) inherent orientation $\theta_i = \theta(z_i + \varepsilon) = 0$.
    }
    \label{fig:OrientationComparison}
\end{figure}

We will determine the set $\{\mu_i\}$ and $\theta_0$ from given phases for each defect in an infinite system with uniform background (see Ref.~\cite{Copar24} for a bounded system).
For the director to be uniform at large distances, we must have $\sum_i\mu_i = 0$. 
Given the orientation $\theta_j = \theta(z_j+\varepsilon)$ for each defect, we may write a linear system of $N+1$ equations determining $\{\mu_i\}$ and $\theta_0$:
\begin{equation} \label{eqn:SpiralChargeSystem}
    \begin{pmatrix}
        A_{11} & \cdots & A_{1N} & 1 \\
        \vdots & & \vdots & \vdots \\
        A_{N1} &\cdots & A_{NN} & 1 \\
        1 & \cdots & 1 & 0        
    \end{pmatrix}
    \begin{pmatrix}
        \mu_1 \\
        \vdots \\
        \mu_N \\
        \theta_0
    \end{pmatrix} =
    \begin{pmatrix}
        \theta_1 - \Theta_1 \\
        \vdots \\
        \theta_N - \Theta_N \\
        0
    \end{pmatrix}
\end{equation}
where
\begin{align}
    A_{ii} &= \log \varepsilon \nonumber \\
    A_{ij} &= \log|z_i - z_j| \label{eqn:SpiralChargeDefs}\\
    \Theta_i &= \sum_{j\neq i} m_j \arg(z_i - z_j). \nonumber
\end{align}
We note that we have assumed the same separation of length scales as in calculating the free energy.
Solving Eq.~\eqref{eqn:SpiralChargeSystem} determines the director in Eq.~\eqref{eqn:MultiDefectSpiralCharge}.
Figure \ref{fig:OrientationComparison} shows two different director configurations for defects in the same location, but with $\mu_i = 0$ for all defects in one and $\theta_i = 0$ for the other.
When $\mu_i=0$ the defects are said to have ``optimal orientation'' since the Oseen-Frank free energy is minimized subject to the constraint of Eq.~\eqref{eqn:2DWindingNumber}.
Nonzero $\mu_i$ is still a local minimum of the Oseen-Frank free energy, but subject to the boundary conditions that $\theta(z_i + \varepsilon) = \theta_i$ and $\sum_i \mu_i = 0$.
As such, the two solutions are technically minimizers of different problems: the former allows  defects to rotate freely so that each $\mu_i = 0$ (so that energy is minimized), while the latter fixes the orientation of each defect, resulting in nonzero $\mu_i$. 

As the number of defects grows, analytically solving Eq.~\eqref{eqn:SpiralChargeSystem} becomes more and more intractable.
For the case of just two defects, the system may be solved relatively painlessly.
Rotating the system axes so that $z_1 = -d/2$ and $z_2 = d/2$ and solving Eq.~\eqref{eqn:SpiralChargeSystem} gives
\begin{align}
    \mu_1 &= \frac{-\theta_1 + \theta_2 + m_2\pi}{2(\log d-\log \varepsilon )} \nonumber\\ 
    \mu_2 &= \frac{\theta_1-\theta_2 - m_2\pi}{2(\log d - \log \varepsilon)} \\
    \theta_0 &= \frac{1}{2}\left(\theta_1 + \theta_2 - m_2\pi\right), \nonumber
\end{align}
which coincides with the two defect solution of Ref.~\cite{tang17} for the appropriate hierarchy of length scales.

\subsection{Near field defect core} \label{subsec:NearFieldDefectCore}
Now that we are equipped with expressions for the director field away from the defect core, we briefly turn to the structure of defect cores.
We now need to know not only the director field, but also the scalar order parameter $S$.
If we substitute $q = Se^{i2\theta}$ into the bulk free energy densities, Eqs.~\eqref{eqn:Landau-deGennes} and ~\eqref{eqn:QOneConstantElastic}, they read $f_b = (A/2) S^2 + (C/4)S^4$ and $f_e = 2K|\partial_z S|^2 + 8K S^2 |\partial_z \theta|^2$.
The corresponding Euler-Lagrange equations govern the equilibrium structure and are given by:
\begin{align}
    4\frac{K}{C}\partial_z\partial_{\zb}S + S\left(1 - S^2 - 16\frac{K}{C}|\partial_z\theta|^2\right) &= 0 \label{eqn:NearCoreS} \\
    \partial_z\partial_{\zb}\theta +\frac{2}{S}\left(\partial_z S \partial_{\zb} \theta + \partial_{\zb} S\partial_z \theta\right) &= 0 \label{eqn:NearCoreTheta}
\end{align}
where $S$ here represents our rescaled scalar order parameter (scaled by $s_0 = \sqrt{-A/C}$).
It is common to introduce a nematic coherence length $\xi = \sqrt{K/C}$ and rescale lengths by $\xi$ so that Eq.~\eqref{eqn:NearCoreS} becomes independent of $K/C$.

For an isolated, stationary defect, it is common to assume that $S = S(|z|)$, that is, a function of radial coordinate only.
This assumption is justified in the one elastic constant approximation since the energy density around the defect solution given in the previous section is isotropic.
We will first assume $\mu = 0$ so $\theta$ is given by Eq.~\eqref{eqn:ComplexSingleDefectDirector}.
Under this assumption, Eq.~\eqref{eqn:NearCoreTheta} is automatically satisfied, while Eq.~\eqref{eqn:NearCoreS} becomes:
\begin{equation} \label{eqn:DefectCoreEquation}
    S'' + \frac{1}{|z|}S' + S\left(1 - S^2 - 4\frac{m^2}{|z|^2}\right) = 0,
\end{equation}
with boundary conditions $S(0) = 0$ and $\lim_{|z|\to\infty}S(|z|) = 1$.
To our knowledge, Eq.~\eqref{eqn:DefectCoreEquation} has no known analytical solution.
Asymptotically, $S(|z|\to0) = |z|^{2|m|}$ and $S(|z|\to\infty) = 1$ satisfy the small $|z|$ and large $|z|$ limits respectively (note that one may ignore the $1 - S^2$ term in the small $|z|$ limit while one may ignore the $m^2/|z|^2$ term in the large $|z|$ limit).

\begin{figure}
\centering
    \includegraphics[width = \columnwidth]{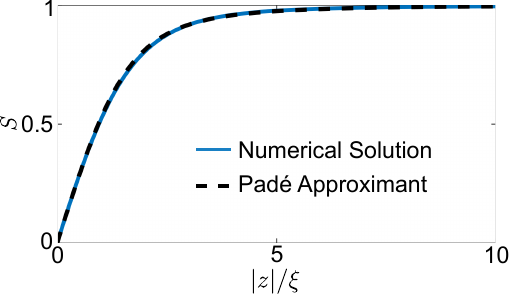}
    \caption{Numerical solution to Eq.~\eqref{eqn:DefectCoreEquation} for the scalar order parameter $S(|z|)$ near a defect core with winding $|m| = 1/2$.
    The dashed black curve is the Pad\'e approximant given in Eq.~\eqref{eqn:PadeApproximateCore}.}
    \label{fig:NearCoreSolution}
\end{figure}

While numerical solutions to Eq.~\eqref{eqn:DefectCoreEquation} are easily obtainable \cite{Ginzburg58,pismen99}, analytical approximations still prove invaluable in obtaining approximations for defect motion and for studying other aspects of defect structure.
The Pad\'e approximant:
\begin{equation} \label{eqn:PadeApproximateCore}
    S^2(|z|) = \frac{|z|^2\left(0.34 + 0.07 |z|^2\right)}{1 + 0.41 |z|^2 + 0.07 |z|^4} 
\end{equation}
gives a very close approximation to the numerical solution, as shown in Fig.~\ref{fig:NearCoreSolution} for an $|m| = 1/2$ disclination.
A cruder approximation, yet one that is very useful for calculation, is the linear core approximation where $S(|z|) = |z|/\varepsilon$ for $|z| < \varepsilon$ and $S(|z|) = 1$ for $|z| > \varepsilon$ (assuming $|m| = 1/2$).
Under this approximation, the complex-valued order parameter near the core takes on a particularly simple form:
\begin{equation}
    q(z,\zb) = \begin{cases}
        z/\varepsilon & m = 1/2,\, |z| < \varepsilon \\
        \zb/\varepsilon & m = -1/2, \, |z| < \varepsilon \\
        \left(z / \zb \right)^{1/2} & |z| > \varepsilon
    \end{cases}
\end{equation}
For well-separated defect cores ($d_{ij} \gg \varepsilon$), we may treat the total order parameter of the system as the product of single defect order parameters (assuming $|q|$ is scaled by $s_0$):
\begin{equation} \label{eqn:OrderParameterProduct}
    q(z,\zb) = \prod_{i} q_i(z,\zb).
\end{equation}
As we will see, this product form will be important for several methods of approximating defect velocities.

Finally, above we assumed that spiral charges are zero.
If we allow nonzero spiral charge, so that $\theta = m\arg(z) + \mu\log(|z|)$, even though $\theta$ is still harmonic Eq.~\eqref{eqn:NearCoreTheta} is no longer identically satisfied since 
\begin{equation}
    \partial_z S \partial_{\zb}\theta + \partial_{\zb} S \partial_z\theta =\frac{\mu S'}{2|z|}.    
\end{equation}
One way to reconcile this could be to relax the assumption that $S = S(|z|)$, that is, to not assume $S$ is isotropically distributed.
However, since the energy density is isotropic, there is no reason for $S$ not to be isotropic.
Another potential solution is to modify $\theta(z,\zb) = m\arg(z) + \tilde{\theta}(|z|)$ and find $\tilde{\theta}$ so that Eqs.~\eqref{eqn:NearCoreS} and \eqref{eqn:NearCoreTheta} are satisfied.
In the region where $S$ is changing most rapidly ($|z|\ll\varepsilon$), a simple solution is $\tilde{\theta}(|z|) = \tilde{\theta}_0$ a constant.
Then Eq.~\eqref{eqn:NearCoreTheta} is still satisfied and Eq.~\eqref{eqn:DefectCoreEquation} describing $S$ is unchanged.
How this near-core constant solution should match up to the far-core spiral charge solution is not obvious, and since we are not aware of any studies that have taken on this problem of incorporating spiral charges into the near field defect solution, we leave it as an open problem for a future study.

\subsection{Two elastic constants}
We will now return to the far-field director solution, and focus on cases that modify Eq.~\eqref{eqn:ManyDefectDirector} for the director angle.
In the previous sections we assumed the Oseen-Frank elastic energy was given by Eq.~\eqref{eqn:1Const2DFrank} with a single elastic constant $K$.
Allowing $K_{11} \neq K_{33}$ the Oseen-Frank free energy reads
\begin{equation} \label{eqn:2Constant2DFrank}
    \mathcal{F}_{OF} = \frac{i}{2}\int_\Gamma \left(2 \bar{K}|\partial_z\theta|^2 - 2 \delta K \Re\left[e^{2i\theta} (\partial_z\theta)^2\right]\right)\, dz d\zb
\end{equation}
where $\bar{K} = (K_{11} + K_{33})/2$ and $\delta K = (K_{11} - K_{33})/2$.
As required, Eq.~\eqref{eqn:2Constant2DFrank} reduces to Eq.~\eqref{eqn:1Const2DFrank} in the limit $K_{11} = K_{33}$.

For the remainder of this section, we will rescale the energy by $2 \bar{K}$ and define the elastic anisotropy $\epsilon = \delta K/\bar{K} = (K_{11} - K_{33})/(K_{11} + K_{33})$ \cite{hudson89}.
The Euler-Lagrange equation corresponding to Eq.~\eqref{eqn:2Constant2DFrank} is
\begin{equation} \label{eqn:AnisoELEquation}
    \partial_z\partial_{\zb} \theta + \epsilon\left\{ \Im\left[e^{2i\theta}\left(\partial_z\theta\right)^2\right] - \Re\left[e^{2i\theta}\partial_z\partial_z\theta\right]\right\} = 0.
\end{equation}
Solutions to the above equation are no longer harmonic, hence solutions are no longer simply the real or imaginary parts of complex analytic functions.
Thus, the electrostatic analogy also breaks down.
Indeed, Eq.~\eqref{eqn:AnisoELEquation} is not even linear, so general solutions are quite difficult to find.

\begin{figure}
\centering
    \includegraphics[width = \columnwidth]{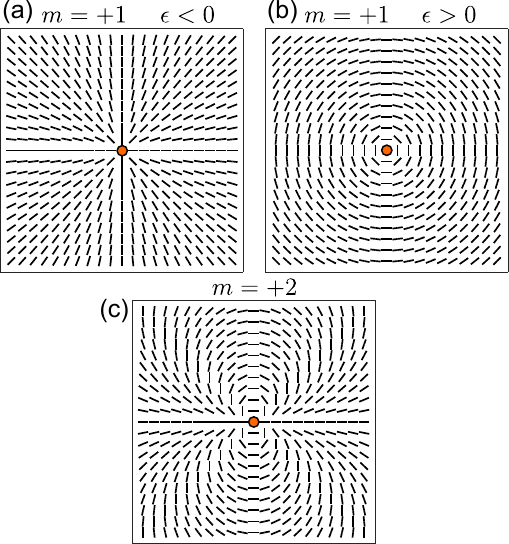}
    \caption{Defect configurations which have the form $\theta(\varphi) = m\varphi + \theta_0$ that also satisfy the anisotropic elastic Euler-Lagrange equation, Eq.~\eqref{eqn:AzimuthalELEquation}.
    (a) $m = +1$ and $\theta_0=0$. This defect configuration has no bend, and thus minimizes the energy (subject to the topological constraint) when $\epsilon < 0$.
    (b) $m = +1$ and $\theta_0 = \pi/2$. This defect configuration has no splay, and thus minimizes the energy when $\epsilon > 0$.
    (c) $m = +2$ also satisfies the Euler-Lagrange equation for any value of $\theta_0$ and $\varepsilon$.}
    \label{fig:SimpleDzyaloshinskiiSolutions}
\end{figure}

While generic solutions of Eq.~\eqref{eqn:AnisoELEquation} are difficult to come by, we may search for solutions that are purely azimuthal by assuming $\theta = \theta(\varphi)$.
In this way, we may find the director angle in the far-field limit for a single defect of charge $m$.
Equation \eqref{eqn:AnisoELEquation} then becomes
\begin{multline} \label{eqn:AzimuthalELEquation}
    \partial_{\varphi}\partial_{\varphi} \theta\left[1 + \epsilon\cos2(\theta-\varphi)\right] \\
    + \epsilon\left(2 - \partial_{\varphi}\theta\right)\partial_{\varphi}\theta\sin 2(\theta -\varphi) = 0.
\end{multline}
An important feature of this equation is that if $\theta$ is a solution for a given $\epsilon$, then $\theta + \pi/2$ is a solution for $\epsilon\to-\epsilon$ since adding $\pi/2$ just changes the sign of the complex exponentials.
Thus, only solutions for $\epsilon \geq 0$ are required. 
For specific solutions, it is immediately apparent that director configurations $\theta = m\varphi + \theta_0$ (i.e. given by Eq.~\eqref{eqn:RealSingleDefectDirector}) with $m = 0$, $+1$, or $+2$ satisfy Eq.~\eqref{eqn:AzimuthalELEquation}. Configurations with $m = +1$ and $m = +2$ are plotted in Fig.~\ref{fig:SimpleDzyaloshinskiiSolutions}.
Configurations with $m=0$ are uniform solutions with no defect, that is, the nematic ground state.
Configurations with $m = +1$ are also obvious from a physical perspective since these defects may be either all splay ($\theta_0 = 0$) or all bend ($\theta_0 = \pi/2$) and so can expel the more expensive type of deformation entirely.
The surprising case is $m = +2$.
The fact that the same $m = +2$ configuration, $\theta = 2\varphi + \theta_0$ is a local minimum of the Oseen-Frank free energy at all $\epsilon$ implies that it is not possible for the director to reduce the net splay or bend energy of a $+2$ defect by any continuous transformation.
If it were possible, one could, for example, reduce the splay region when $\epsilon = 1$ (bend constant zero) and therefore reduce the total free energy.
Another way to say this is the configuration $\theta = 2\varphi + \theta_0$ has minimal bend \textit{and} splay energies subject to  the topology $m = +2$, for arbitrarily small measuring loops.
Of course, it is likely that two separated $m = +1$ defects (so that the total winding is still $+2$) cost less energy, so that a $m = +2$ defect would be unstable to splitting (and likewise for the $m=+1$ defects into two $m=+1/2$ defects).

We now focus on the case of $m = \pm 1/2$, which still cost the least amount of energy individually and are the most commonly seen winding in experiments and simulations \cite{deGennes75}.
An implicit integral solution for $ -1 \le \epsilon \le  1$ was first given by Dzyaloshinskii \cite{dzyaloshinskii70} (though it seems online versions of this article have been lost to time).
The Dzyaloshinskii solution is \cite{hudson89}
\begin{align}
    \varphi &= p\int_0^{\theta(\varphi)-\varphi}\sqrt{\frac{1 + \epsilon\cos 2x}{1 + p^2 \epsilon \cos2x}}\, dx \label{eqn:DzyaloshinskiiSolution} \\
    \pi &= (m - 1) p \int_0^\pi \sqrt{\frac{1 + \epsilon\cos 2x}{1 + p^2 \epsilon \cos2x}}\, dx. \nonumber
\end{align}
Here, $p$ is a constant determined by the second equation, which is essentially a boundary condition set by the periodicity of azimuthal solutions: the director must rotate by $2\pi m$ over a loop surrounding the defect core.
Note that Eq.~\eqref{eqn:DzyaloshinskiiSolution} gives Eq.~\eqref{eqn:RealSingleDefectDirector} when $\epsilon = 0$.

\begin{figure}
\centering
    \includegraphics[width = \columnwidth]{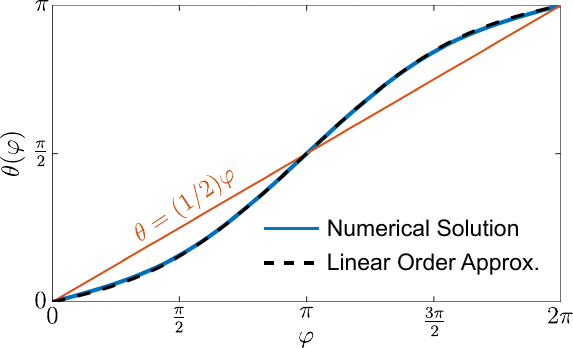}
    \caption{The director angle $\theta(\varphi)$ as a function of azimuth $\varphi$ for elastic anisotropy $\epsilon=0.4$.
    The solid blue line shows the numerical solution to Eq.~\eqref{eqn:AzimuthalELEquation}, while the dashed black line shows the linear order approximation given in Eq.~\eqref{eqn:LinearOrderEpsilonTheta}.
    The solid orange line shows the isotropic solution $\theta(\varphi) = (1/2)\varphi$.} 
    \label{fig:AnisotropicApproximation}
\end{figure}

For arbitrary elastic anisotropy $\epsilon$, there is no known closed form solution, and one must again turn to numerics.
There are special cases where an analytical approximation or full solution may be obtained, however.
First, one may expand either the differential equation~\eqref{eqn:AzimuthalELEquation} or the integral solution~\eqref{eqn:DzyaloshinskiiSolution} in powers of $\epsilon$ to establish linear order solutions (i.e. for $|\epsilon| \ll 1$) \cite{zhou17,Myers24}.
To first order in $\epsilon$ the solution is:
\begin{equation} \label{eqn:LinearOrderEpsilonTheta}
    \theta(\varphi) = m\varphi - \epsilon \frac{m(2-m)}{4(1-m)^2}\sin\left[2(1-m)\varphi\right] + \theta_0
\end{equation}
which is plotted in Fig.~\ref{fig:AnisotropicApproximation} for $\epsilon = 0.4$ compared to the numerical solution of Eq.~\eqref{eqn:DzyaloshinskiiSolution}.
One may continue the perturbative solution to any finite order in $\epsilon$, though comparison with the numerical solution tells us that Eq.~\eqref{eqn:LinearOrderEpsilonTheta} gives a good approximation to the solution up to $\epsilon \approx 0.6$.

\begin{figure}
\centering
    \includegraphics[width = \columnwidth]{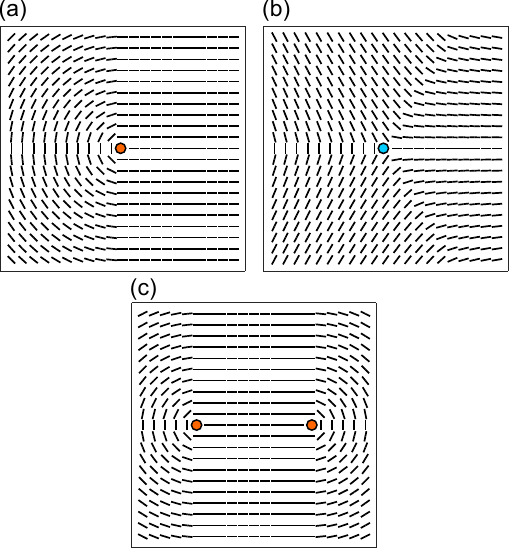}
    \caption{Director configurations for $m=\pm1/2$ defects with extreme anisotropy $\epsilon=1$.
    (a) A $+1/2$ defect which expels all splay given by Eq.~\eqref{eqn:AnisotropicPlusHalf}.
    (b) A $-1/2$ defect which expels as much splay as possible given by Eq.~\eqref{eqn:AnisotropicMinusHalf}.
    (c) Two $+1/2$ defects which expel all splay.
    This configuration is given by joining two halves of a $+1$ defect with a uniform director configuration.} 
    \label{fig:AnisotropicConfigurations}
\end{figure}

The other special case is extreme anisotropy, $\epsilon = \pm1$.
Since this case requires an elastic constant going to zero, the physical relevance of these solutions in pure nematics is questionable, though they may find use for defects in other liquid crystalline phases such as smectics or cholesterics which tend to expel certain types of nematic distortion \cite{deGennes75,Wang06}.
We consider the case where $\epsilon = +1$, i.e. $K_3 \to 0$.
A minimal energy configuration for an $m = +1/2$ defect cannot be found directly from the Dzyaloshinskii solution, Eq.~\eqref{eqn:DzyaloshinskiiSolution}, without careful regularization \cite{Klein2026Signatures}.
It can, however, be intuited from the fact that the system wants to expel splay:
since both the uniform state and $+1$ defects have configurations without splay, we may stitch the two configurations together, while preserving the topological constraint of $m = +1/2$, as shown in Fig.~\ref{fig:AnisotropicConfigurations}(a).
Analytically, the angle of the director is given by the piecewise function
\begin{equation} \label{eqn:AnisotropicPlusHalf}
    \theta(\varphi) = \begin{cases}
        \varphi + \frac{\pi}{2} & \frac{\pi}{2} < \varphi - \theta_0 < \frac{3\pi}{2} \\
        \theta_0 & \text{else}
    \end{cases}
\end{equation}
where $\theta_0$ is the angle offset which gives the orientation of the defect.
We note that for a charge $m = +3/2$ defect, a similar construction is obtainable \cite{Klein2026Signatures}.
For $m < 0$ and $m > 2$, Eq.~\eqref{eqn:DzyaloshinskiiSolution} may be solved exactly when $\epsilon = \pm1$.
While a general solution exists \cite{Klein2026Signatures}, here we give the $m = -1/2$, $\epsilon = +1$ solution, plotted in Fig.~\ref{fig:AnisotropicConfigurations}(b).
The director angle is
\begin{align}
    \theta(\varphi) &= \varphi - \arctan\left[\frac{\sin \left(\varphi + 2n\pi/3\right)}{\sqrt{\cos^2\left(\varphi + 2n\pi/3\right) - 1/4}}\right]+n\pi \label{eqn:AnisotropicMinusHalf} \\
    n &= \left\lfloor\frac{1}{2} - \frac{3}{2\pi}\varphi\right\rfloor \nonumber
\end{align}
where $\lfloor\cdot\rfloor$ is the floor function.

For configurations with multiple defects, the defect solutions cannot be superimposed as they can in the isotropic elasticity limit since the governing differential equation, Eq.~\eqref{eqn:AnisoELEquation}, is nonlinear.
Only in the case of two $m = +1/2$ defects in the extreme anisotropy scenario ($\epsilon = \pm 1$) can one find a director field by splitting apart an $m = +1$ defect and filling the center with a uniform director field [see Fig.~\ref{fig:AnisotropicConfigurations}(c)].
In the case of small anisotropy, the equations may be solved perturbatively, though to our knowledge investigations in this direction have still relied on numerics \cite{Myers24,Houston25}.

We briefly mention that the near core order parameter also changes as a result of anisotropic elasticity.
The elastic anisotropy induces anisotropy in the scalar order parameter as well, so the assumption $S = S(|z|)$ no longer holds and we must either directly solve for the order parameter $q$, or perturbatively solve for $S$ and $\theta$ near the core.
The full solution for $q$ has been investigated numerically \cite{zhou17,schimming20b}, while a perturbative solution has been explored recently in the context of defect dynamics \cite{Montagna26}.

Finally, we mention that the introduction of spiral charges to defects in systems with unequal elastic constants has been, to our knowledge, completely unexplored analytically or numerically.
It is not even known whether similar solutions (i.e. with spiral charges) should exist due to the nonlinearity of Eq.~\eqref{eqn:AnisoELEquation}.

For the remainder of this review, unless otherwise specified, we will return to assuming a single elastic constant free energy.

\subsection{Dirichlet boundary conditions}
Thus far, we have assumed the nematic lives in an infinite two-dimensional domain in which the conditions at infinity are mostly ignored.
That situation is, of course, unphysical, yet it is still a very useful approximation for a system of dilute topological defects which are far from each other and far from the domain boundaries.
However, boundary conditions do indeed alter the structure (and dynamics) of topological defects in nematics.
Here we will consider two types of boundary conditions: Dirichlet (where the director angle $\theta$ is prescribed on the boundary) and periodic (where the director angle must match on opposite sides of the domain). Dirichlet conditions, where the director is held fixed at the boundary, are a ``strong anchoring'' limit in which the anchoring extrapolation length $K/W \rightarrow 0$ (here $W$ is the strength of the anchoring in the surface energy density $f_S$).  
One might also consider ``weak anchoring'' where the surface free energy density competes with the other contributions to the free energy.
We do not consider weak anchoring here, but mention that several methods have been established to yield approximate, analytic solutions to the director field in these cases \cite{Lewis17,paparini21,paparini21b,Chandler23,Chandler24}.

\begin{figure}
\centering
    \includegraphics[width = \columnwidth]{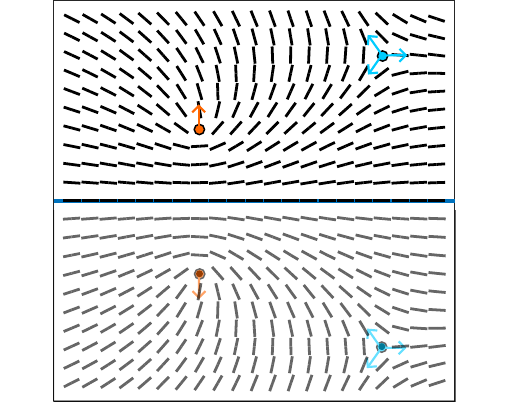}
    \caption{Defect configuration with Dirichlet boundary at $\Im[z] = 0$ of $\Theta_0 = 0$.
    Image charges are placed at locations $\zb_i$ with equal winding and opposite spiral charge so that the director field is mirrored across $\Im[z] = 0$.} 
    \label{fig:HalfPlaneBoundary}
\end{figure}

The simplest domain with boundary to consider is the half-infinite plane.
For concreteness, let us take the upper half-plane $\Gamma = \{z \in\mathbb{C}\,|\, \Im[z] \geq 0\}$ whose boundary is the real line $\partial \Gamma =\{z\,|\,\Im[z] = 0\}\subset\mathbb{C}$.
We then take as boundary condition $\theta(\Im[z] = 0) = \Theta_0$, a constant.
For a defect configuration with defects of charge $m_i$ located at $z_i$, the solution is given by adding image defects across the boundary of the same winding but opposite spiral charge \cite{deGennes75}:
\begin{multline} \label{eqn:SingleBoundaryDirector}
    \theta(z,\zb) = \sum_i \left\{m_i\left[\arg(z - z_i) + \arg(z - \zb_i)\right] \right.\\ 
    + \left.\mu_i\left[\log|z-z_i| - \log|z - \zb_i|\right]\right\} + \Theta_0.
\end{multline}
An example configuration is plotted in Fig.~\ref{fig:HalfPlaneBoundary}.

Taking the spiral charges $\mu_i=0$ in Eq.~\eqref{eqn:SingleBoundaryDirector}, we see a striking similarity (or dissimilarity) to electrostatic charges.
In our analogy to electrostatics, fixing the director angle at a boundary is like setting the potential at the surface of a conductor.
However, recall from elementary electrostatics that the image charge solution adds \textit{oppositely} charged particles as images.
Hence, for bounded systems, the analogy breaks down, though in a symmetric way for Dirichlet conditions.
We may have anticipated this from the mathematical form of the analogy, namely that the electrostatic potential and the director angle are the real and imaginary parts of a complex analytic function: there are no analytic functions that satisfy both requirements of a fixed Dirichlet boundary condition and a nonzero number of poles (which are the charges or defects).
It is only possible for either the real or imaginary parts of such a function to be fixed along a curve in the complex plane.

While Eq.~\eqref{eqn:SingleBoundaryDirector} applies to the case of an unphysical domain, it is actually a very important director configuration.
This is because it may be transformed to a physically relevant domain via conformal mapping.
A conformal map at a point in the complex plane is simply a complex analytic function with nonzero derivative at that point (if all points have nonzero derivative then it is a conformal map over the entirety of the domain)\cite{Arfken05}.
A standard result in complex analysis tells us that a harmonic function in one domain remains harmonic in another domain that is transformed via a conformal map.
Thus, if we have a bounded domain $\Gamma$, and know the boundary condition on $\partial\Gamma$ we may first transform this domain back to the half plane, then use image charges to solve for the director field on the half plane, then transform back to get our director field everywhere in $\Gamma$. 
Depending on the boundary condition, it may or may not be difficult to solve the system on the half plane, but it will likely be the simplest domain that solves the equations.

We illustrate the above with an explicit, important example.
Let our domain of interest be the unit disc, $\tilde{\Gamma} = \{w\in \mathbb{C}\,|\, |w| \leq 1\}$.
Of course, the boundary $\partial \tilde{\Gamma}$ is the unit circle defined by $|w| = 1$.
We take as the boundary condition a fixed defect profile, $\tilde{\theta}(|w|=1,\varphi) = M\varphi + \theta_0$, so that $M$ is the total winding in the domain.
We must have that $\sum_i m_i = M$ in the domain, but we are otherwise free to search for director configurations with defects with arbitrary windings and orientations at arbitrary points in the domain, exactly the problem we have been focusing on above in an infinite domain.
The simplest way to solve this problem is to use a conformal map to the upper half plane, which is given by the M\"obius transformation (a conformal map given by a ratio of linear functions):
\begin{align}
    z &= \frac{w + i}{iw + 1}. \label{eqn:CircleConformalInverse} 
\end{align}
so that $z$ is the coordinate in the upper half plane; the inverse map is 
\begin{align}
        w &= \frac{z - i}{1-iz} .\label{eqn:CircleConformal}  
\end{align} 
This transformation sends the unit circle $|w| = 1$ to the real  line $\Im[z] = 0$,  the interior of the disc to the upper half plane, and the exterior of the disc  to the lower half plane.
The boundary condition is transformed to
\begin{equation} \label{eqn:CircleBoundaryCondition}
    \theta(\Im[z] = 0) = -2M\left[\arg(\Re[z] + i) - \pi/2\right] + \theta_0
\end{equation}
which can be shown by using the properties $\arg(z_1z_2) = \arg(z_1) + \arg(z_2)$ and $\arg(z) = -\arg(\zb)$.

\begin{figure*}
\centering
    \includegraphics[width = \textwidth]{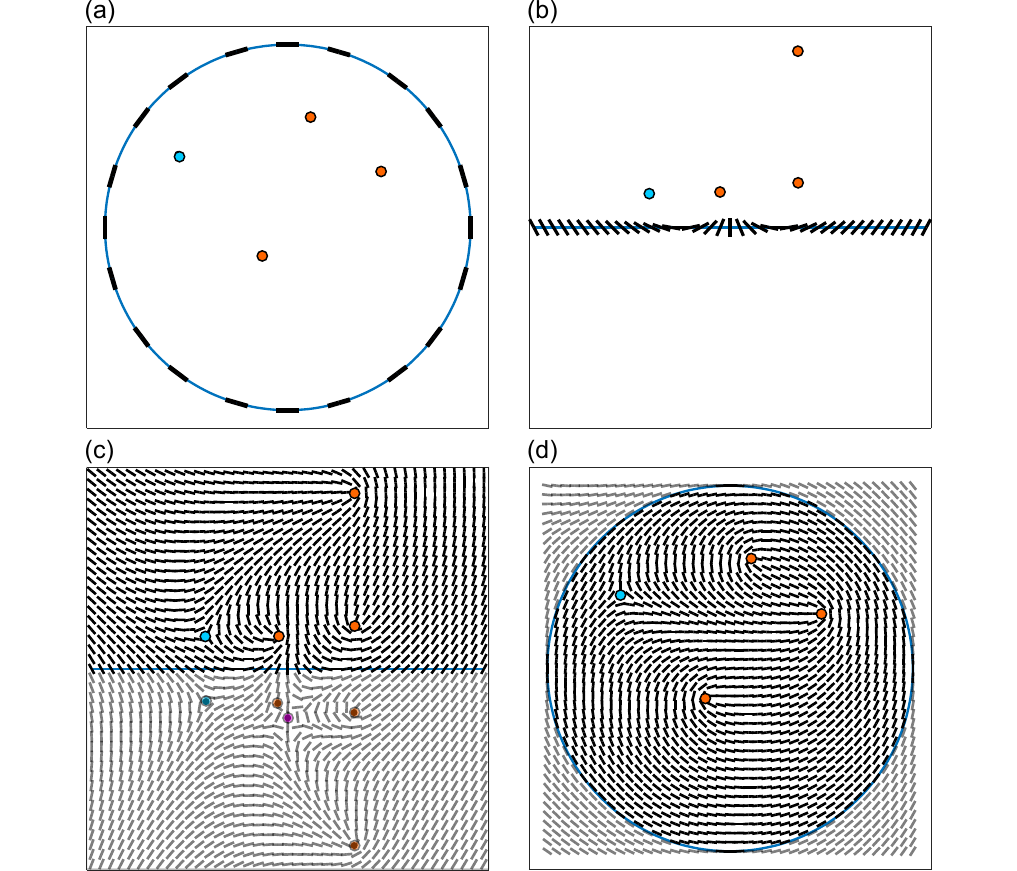}
    \caption{Finding the director field in a disc for arbitrary defect placement.
    (a) A boundary condition where $\theta = \varphi +\pi/2$ is set and four defects are placed, three with winding $m=+1/2$ and one with winding $m=-1/2$. 
    Note that since the overall winding is $M =+1$ the sum of all defect windings is also $\sum_i m_i = +1$.
    (b) The defect positions and boundary conditions are mapped to the half plane using the M\"obius transformation Eq.~\eqref{eqn:CircleConformalInverse}.
    (c) The director field is obtained on the half plane using virtual image defects.
    The purple virtual defect located at $z = -i$ has winding $-2M = -2$ to set the correct boundary condition.
    (d) The director field from the half plane is transformed back into the original disc system using Eq.~\eqref{eqn:CircleConformalInverse}.}
    \label{fig:ConformalMappingExample}
\end{figure*}

If $M = 0$, we may use Eq.~\eqref{eqn:SingleBoundaryDirector} with $\Theta_0 = \theta_0$ and defects located at $z(w_i)$, then substitute Eq.~\eqref{eqn:CircleConformalInverse} to get the director field in terms of the disc coordinates.
If $M \neq 0$, we may still use Eq.~\eqref{eqn:SingleBoundaryDirector}, but we must add to it a defect distribution that satisfies the boundary condition (we may add solutions because the Euler-Lagrange equation is linear).
To match the boundary condition of Eq.~\eqref{eqn:CircleBoundaryCondition} we must add an additional virtual defect (with no real image in the upper half plane) at $z = -i$ of winding $-2M$ as well as a global phase of $\pi M + \theta_0$.
Conceptually, we note that $z = -i$ is mapped to $w = \infty$ in the disc system.
Adding a virtual defect at infinity is the price we pay for mapping an infinite domain to a bounded domain with nonzero total winding.
Converting to coordinates $w$ and performing algebraic simplifications, our director field in the unit disc is given by
\begin{multline} \label{eqn:BoundedCircleDirector}
    \tilde\theta(w,\bar w) = \\
    \sum_i\left\{ m_i\left[\arg(w - w_i) +\arg\left(w - \frac{1}{\bar w_i}\right) - \arg\left(w_i\right) \right] \right. \\
    \left. + \mu_i \left[\log |w - w_i| - \log \left|w - \frac{1}{\bar w_i}\right| - \log|w_i|\right]\right\} + \tilde\Theta_0
\end{multline}
where
\begin{equation}
    \tilde\Theta_0 = \theta_0 - \frac{\pi}{2}\sum_i m_i
\end{equation}
sets the director orientation at the boundary of the unit circle.
We have written Eq.~\eqref{eqn:BoundedCircleDirector} in a way that highlights the fact that the virtual defects are now placed at reciprocal distances from the real defects.
Note that the terms which add constant pieces, $\sum_i m_i\arg(w_i)$ and $\sum_i \mu_i \log|w_i|$ may also be absorbed into the definition of $\tilde\Theta_0$.
We show a visual example of this process in Fig.~\ref{fig:ConformalMappingExample} where we have set $\mu_i =0$ for simplicity.
The case where orientations of defects are fixed may be solved by first transforming the orientations then computing spiral charges for the half plane system or, alternatively, by finding the spiral charges directly from the director field that has been transformed back by adding the appropriate logarithmic terms.

The above process works for any simply connected domain (as long as one can solve the boundary value problem on the half plane).
The case in which the domain is not simply connected is more difficult to handle.
Recent work has shown that exact solutions for the equilibrium director field may be found for multiple inclusions.
For example, Chandler and Spagnolie \cite{Chandler24} find analytical solutions for domains with two disks or triangles as inclusions, and use these solutions to compute the effective forces between the inclusions.
While these calculations are similar in process (though more difficult) to the example shown above, they build upon recent advancements in general conformal mapping techniques for multiply connected domains \cite{Crowdy06,Crowdy20}.
However, to our knowledge, no general configurations for arbitrary defects are known analytically except in particular geometries such as the annulus \cite{Lewis17}.

\subsection{Periodic boundary conditions}
Apart from Dirichlet conditions, another common boundary condition (particularly in numerical studies) is periodic.
For periodic boundary conditions, there is a strict topological constraint, $\sum_i m_i = 0$.
This is due to the Poincar\'e-Hopf theorem, which states that the total winding of a line field on a closed surface must be equal to the surface's Euler characteristic.
A periodic domain is topologically equivalent to the torus, which has Euler characteristic $0$.
For simplicity we will here consider a square domain of size $2L$ (the case of an arbitrary parallelogram domain can be handled with minor adjustments \cite{Schimming25b}).

\begin{figure}
\centering
    \includegraphics[width = \columnwidth]{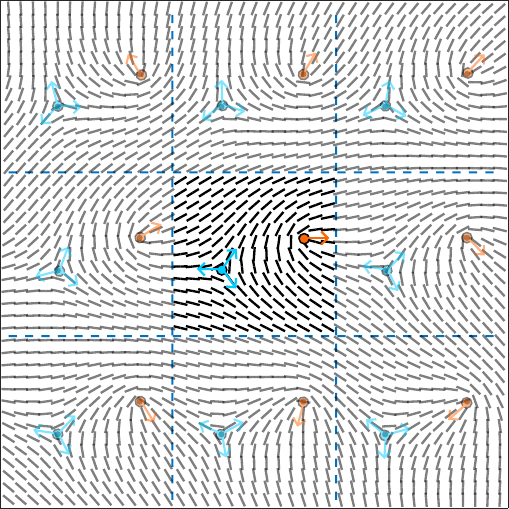}
    \caption{Director configuration for two defects and their periodic images computed from $\theta(z,\zb) = \sum_i m_i \sum_{n,\ell} \arg\left[z - (z_i + 2n L + 2i\ell L)\right]$.
    The blue dashed lines separate the center cell from the periodic images.
    Note that the defect orientations are not maintained in the images.} 
    \label{fig:PeriodicMisalignment}
\end{figure}

One might be tempted to use image charges again to solve the problem: take $\theta(z,\zb) = \sum_i m_i \sum_{n,\ell}\arg\left[z - (z_i + 2 n L +2 i \ell L)\right]$ where $n$ and $\ell$ are integers.
However, as one can easily show numerically, this configuration does not converge to be periodic!
Why is this?
It becomes apparent if one plots the images themselves (Fig.~\ref{fig:PeriodicMisalignment}).
The orientations of the image defects do not align with those of the center cell (our actual domain).
Thus, to account for the inconsistent orientations, we must add spiral charge terms to our images.
Solving for the correct spiral charges involves solving an infinite linear system, although numerically we find that the system converges quickly in the number of rings of images added (i.e., one may introduce a cutoff to $|n|$ and $|\ell|$).
While this is one way of accounting for periodic boundaries, there is a more elegant, closed form solution.

As we did for the infinite system, we can write the director angle as the imaginary part of a complex analytic function with poles as the defects.
We now search for a complex analytic function that is doubly periodic (periodic along the real and imaginary axes).
However, there are no doubly periodic analytic functions with arbitrary poles, even if the total winding is zero.
The best we can do is find a function whose imaginary part (but perhaps not the real part) is doubly periodic.
To this end, we introduce the complex Weierstrass $\zeta$ and Weierstrass $\sigma$ functions, defined on our domain \cite{Whittaker_Watson_1996,NIST:DLMF,Hill21,Schimming25b}:
\begin{align}
    \zeta(z) &= \frac{1}{z} + \sum_{n,\ell\neq0}\left[\frac{1}{z - (2 n L + 2i \ell L)}\right. \nonumber \\
   & \qquad\left. +\frac{1}{2 n L + 2i\ell L} + \frac{z}{(2 n L + 2i\ell L)^2}\right] \\
    \partial_z \text{Log} \left(\sigma\right) &= \zeta(z)
\end{align}
where the second equation implicitly defines $\sigma(z)$.
Both $\zeta$ and $\sigma$, though analytic, are only quasi-periodic (that is, $\zeta(z + 2 nL + 2imL) = \zeta(z) + C$ for some constant $C$).
By adding linear functions we may find analytic functions with fully periodic real or imaginary parts.
Using intuition from the infinite domain, one may reason that the derivative of the director angle in the periodic domain should behave similar to the $\zeta$ function above.
Starting from there, it can be shown (see \cite{Schimming25b}) that the fully periodic director angle is
\begin{multline} \label{eqn:PeriodicDirector}
    \theta(z,\zb) = \sum_i m_i\left\{\arg\left[\sigma(z-z_i)\right] - \frac{\pi}{4 L^2}\Im\left[\zb_i z\right]\right\} \\
    + \frac{\pi}{L}\Im\left[(i n_1 + n_2)z\right] + \theta_0
\end{multline}
where $n_1$ and $n_2$ are integers and the total winding $\sum_i m_i =0$.
The second to last term in Eq.~\eqref{eqn:PeriodicDirector} encodes the possibility of having overall winding in either the $x$ or $y$ direction, and should be included even if no defects are present \cite{Schimming25b}.

\begin{figure}
\centering
    \includegraphics[width = \columnwidth]{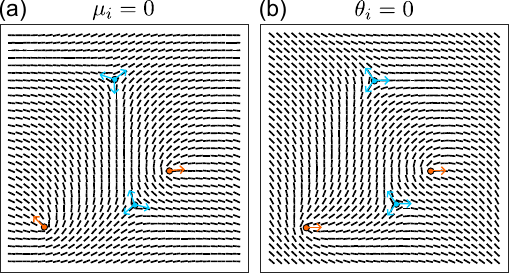}
    \caption{Comparison of two periodic nematic configurations with defects in the same location given by Eq.~\eqref{eqn:FullPeriodicDirector}.
    (a) Spiral charges $\mu_i = 0$ (and global phase $\theta_0 = 0$). 
    (b) Inherent orientations $\theta_i = \theta(z_i + \varepsilon) = 0$.
    }
    \label{fig:PeriodicConfigurations}
\end{figure}

The director configuration given above does not allow control over the local orientation of the defects.
To add local orientations as individual degrees of freedom, we may extend the logic above for an infinite system, and add terms corresponding to a fully periodic real part of an analytic function.
Putting everything together we write the director angle as
\begin{multline} \label{eqn:FullPeriodicDirector}
    \theta(z,\zb) = \sum_i m_i\left\{\arg\left[\sigma(z-z_i)\right] - \frac{\pi}{4 L^2}\Im\left[\zb_i z\right]\right\} \\
    +\sum_i \mu_i\left\{\log|\sigma(z - z_i)| + \frac{\pi}{4L^2}\Re\left[\zb_i z\right]\right\} \\
    + \frac{\pi}{L}\Im\left[(i n_1 + n_2)z\right] + \theta_0
\end{multline}
where the $\mu_i$ and $\theta_0$ are determined from a linear system of equations by matching each individual defect orientation $\theta_i$ with the director at $z = z_i + \varepsilon$ (where $\varepsilon \ll L$ is the defect core size) and ensuring that $\sum_i \mu_i=0$.
We plot an example configuration with optimal orientations ($\mu_i = 0$) and the same configuration with all orientations $\theta_i = 0$ in Fig.~\ref{fig:PeriodicConfigurations} to compare.
In this periodic square domain, Eq.~\eqref{eqn:FullPeriodicDirector} fully determines the far field director given the degrees of freedom $m_i$, $z_i$, and $\theta_i$.

\section{Defect Dynamics} \label{sec:DefectDynamics}

In the previous sections, we focused on what the director field is given a set of static, unmoving defects and boundary conditions.
What happens if we stop holding the defects fixed in place?
Equivalently, we ask: given a set of defect positions, orientations, and boundary conditions, what are the velocities and subsequent motion of all of the defects in the system?

\begin{figure*}
\centering
    \includegraphics[width = \textwidth]{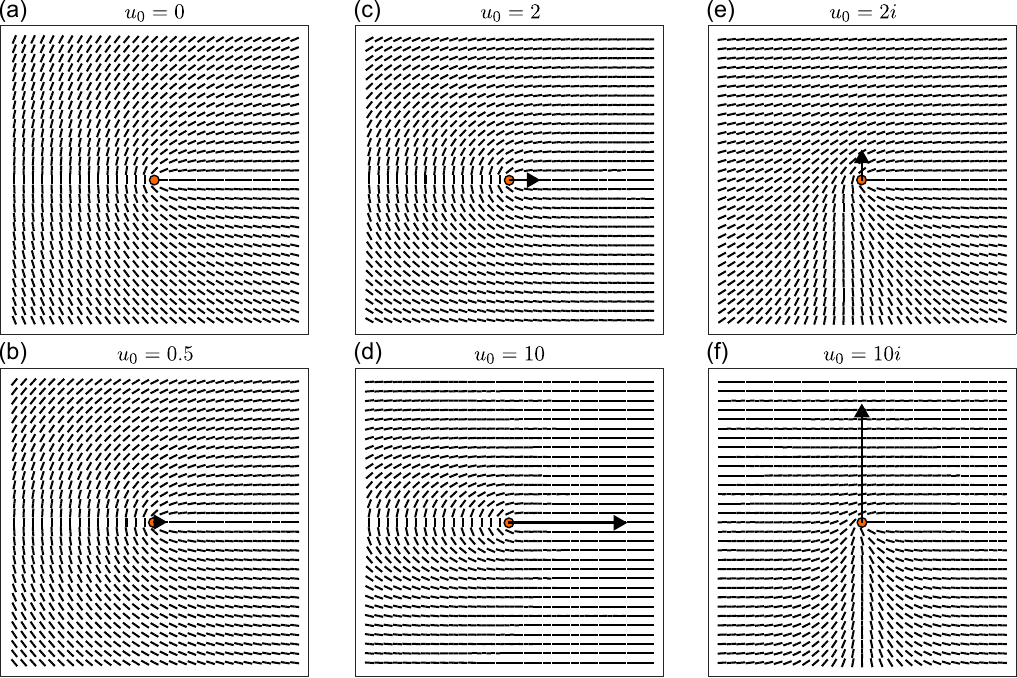}
    \caption{Steady-state director field for an $m=+1/2$ defect moving with velocity $u_0 = $ (a) $0$, (b) $0.5$, (c) $2$, (d) $10$, (e) $2i$, or (f) $10i$ given by Eq.~\eqref{eqn:MovingDefectDirector}.}
    \label{fig:MovingDefectDirector}
\end{figure*}

\subsection{Quasi-static versus Velocity Distorted Configurations}

Before reviewing methods to tackle this question, there is a separate question related to the previous sections we must first reckon with:
the director field for a static defect and the director field for a defect in motion are not necessarily the same!
Indeed, the equations governing the director field depend on the velocity of defect, inducing anisotropy from both the speed and direction of motion \cite{Denniston96,Radzihovsky15}.
Here we will discuss two approximations for the director field of a moving defect.

The first approximation does not change anything from the previous section.
Instead, one assumes that the defect's motion is  sufficiently slow so that the director configuration always has time to relax to a free energy minimum before the defect moves far enough to appreciably change that ``quasi-static'' director field. This is the limit of low Ericksen number $\mathcal{E} = |u_0| \gamma R / K \ll 1$, where $|u_0|$ is the defect's speed and $R$ is the spatial extent of the system.  
For a single defect moving at constant velocity $u_0 = u_{0,x} + iu_{0,y}$, the director is given by
\begin{equation}
    \theta(z,\zb,t) = \theta_{\text{static}}(z,\zb\,|\,z_0\to z_0 + u_0t)
\end{equation}
where the subscript $\text{static}$ indicates that the $\theta$ from the previously described static configurations is used and we replace the defect position $z_0$ with $z_0 + u_0t$.

For faster defects, one must take into account the effect of the defect velocity on the director field.
For a defect moving at constant velocity $u_0$, we may do this by boosting to the co-moving frame of the defect, shifting our coordinates $z' = z - u_0t$.
We then seek a steady state director field which appears constant in this moving frame. The steady state director angle is given by 
\begin{equation} \label{eqn:MovingDefectELEquation}
    4\frac{K}{\gamma}\partial_{z'}\partial_{\zb'}\theta + \left[u_0\partial_{z'}\theta + \bar{u}_0\partial_{\zb'}\theta\right] = 0.
\end{equation}
Equation \eqref{eqn:MovingDefectELEquation} may be solved by decomposing $\theta$ into singular and non-singular components and then using a Green's function \cite{Pismen90,Denniston96,Romano23,Romano24} or Fourier transform \cite{Radzihovsky15}.
Using a Green's function (see e.g. \cite{Denniston96}) the steady state director angle is given by (in real-valued coordinates):
\begin{align}
    \theta(r,\varphi) &= \theta_0 + \int_0^\varphi \left(\nabla G +\frac{\gamma}{K}\mathbf{u}_0 G\right)\cdot \mathbf{r}' \, d\varphi' \nonumber \\
    G(r,\varphi') &= -m \exp\left[-\frac{\gamma}{2K} \mathbf{u}_0\cdot\mathbf{r}'\right]K_0\left(\frac{\gamma}{2K}|\mathbf{u}_0| r\right) \label{eqn:MovingDefectDirector}
\end{align}
where the $'$ only applies to the angular coordinate and $K_0$ is the modified Bessel function of the second kind.
The resulting director field for an $m= +1/2$ winding defect is compared with the quasi-static solution for various defect speeds in Fig.~\ref{fig:MovingDefectDirector}.
We note that the assumption of a steady state director field assumes we are looking at a defect that has been moving forever.
To account for defects that have not been moving forever, one must solve the time dependent relaxation equation, in which case the motion of the defect will depend on the history of its trajectory \cite{Romano23,Romano24}. 

Finally, it will often be convenient to decompose the director field into the singular part corresponding to the defect of interest, and an ``external'' non-singular part corresponding to the contributions of other defects, boundary conditions, or hydrodynamic effects.
Such a decomposition allows us to extract the effects of ``external'' sources on the dynamics of a particular defect.
That is, if the external contribution is zero, we do not expect a defect to begin moving (the defect cannot exert a force on itself).
Thus it is common to write the defect director field and order parameter in the following manner:
\begin{align}
    n(z,\zb,t) &= \left[e^{i\theta_{d}(z,\zb,t)}\right]e^{i\theta_{\text{ext}}(z,\zb,t)} \label{eqn:DirectorDecomp} \\
    q(z,\zb,t) &= \left[S(z,\zb,t)e^{2i\theta_{d}(z,\zb,t)}\right]e^{2i\theta_{\text{ext}}(z,\zb,t)} \label{eqn:OrderParameterDecomp}
\end{align}
where $\theta_d$ is the singular part of the director configuration near the $j$th defect (due to the $j$th defect), while $\theta_{\text{ext}}$ is the nonsingular part of the director configuration near the $j$th defect (due to other defects or boundary conditions, for example).
Decomposing the director or order parameter in this way allows us to apply various approximations to the region near the $j$th defect, as we will see below.
Additionally, it is conceptually beneficial to think of $\theta_d$ as the director configuration for a single defect in an infinite domain (either static or moving at a constant velocity).
It is clear, then, that $\theta_d$ should not contribute to the force on the defect of interest, and thus rotations of the director (given by $\theta_{\text{ext}}$) are responsible for forces on defects and their subsequent motion.

We now discuss some of the more common methods used to derive the velocities of defects, given a particular configuration.
We note that the following subsections do not represent a comprehensive set of methods that achieve this purpose, though they are all commonly used in the recent literature.

\subsection{Force Balance} \label{subsec:ForceBalance}
Some of the earliest efforts to understand disclination motion treated the defects as overdamped point particles.
The equation of motion for a defect is then given by the force balance condition (where the force is given by the complex number $F = F_x + i F_y$)
\begin{equation} \label{eqn:ForceBalance}
    F_{\text{elastic}} + F_{\text{drag}} + F_{\text{hydro}}=0
\end{equation}
where $F_{\text{elastic}}$ is the elastic force on a defect, $F_{\text{drag}}$ is the dissipative drag force on a moving defect, and $F_{\text{hydro}}$ are forces derived from flow fields interacting with the defect.
We will ignore the effects of flow on defects until Sec.~\ref{subsec:Hydro}, and assume $F_{\text{hydro}}=0$ for the rest of this subsection.

From Eq.~\eqref{eqn:ForceBalance}, the problem boils down to finding the forces for a given set of defects and boundary conditions.
The elastic force on the $j$th defect is simply given by the derivative of the free energy with respect to the $j$th defect coordinate \cite{Dafermos70,deGennes75}:
\begin{equation} \label{eqn:DefectElasticForce}
    F_{\text{elastic},j} = -\partial_{\zb_j}\mathcal{F}(\{z_k\})
\end{equation}
where we have written the free energy explicitly as a function of defect positions.
This can be manipulated further by moving the derivative inside the integral over the spatial domain, using Stokes' theorem, and invoking the director decomposition of Eq.~\eqref{eqn:DirectorDecomp} (detailed calculations may be found in e.g. Refs.~\cite{Dafermos70,Romano24}).
If one further assumes a quasi-static director field with vanishing spiral charges $\mu_k =0$ and a one elastic constant free energy, the elastic force may written as
\begin{equation} \label{eqn:ForceExternalField}
    F_{\text{elastic},j} = -2\pi i K m_j \left.\partial_{\zb}\theta_{\text{ext}}\right|_{z=z_j}
\end{equation}
so that the force is explicitly due to the ``external'' part of the director field.
For a system of $N$ defects, we may substitute $\theta_{\text{ext}} = \sum_{k\neq j}^N m_k\arg(z - z_k) + \theta_0$, giving the elastic force
\begin{equation}
    F_{\text{elastic},j} = \pi K \sum_{k\neq j}^N \frac{m_j m_k}{\zb_j - \zb_k}
\end{equation}
which is attractive between opposite sign winding and repulsive between like sign winding.
Note that this force assumes a quasi-static director field, as the energy was calculated from the equilibrium director configuration.

One method of computing the drag force assumes it to be proportional to the defect velocity
\begin{equation}
    F_{\text{drag}} = -\xi u_0
\end{equation}
where $\xi$ is a phenomenological friction coefficient.
The dependence of the friction coefficient on material parameters was first calculated by Imura and Okano \cite{Imura73}, where they argue the energy dissipation rate of the director, $\int\gamma (d \theta /d t)^2\,dA$ (where $\gamma$ is the rotational viscosity, see Eq.~\eqref{eqn:BerisEdwards}), must be equal to the energy dissipation rate of the moving defect, $\xi |u_0|^2$.
Under the quasi-static approximation, the time dependence of the director field for a defect moving at velocity $u_0$ is
\begin{equation}
    \theta(z,\zb,t) = m\arg(z + u_0t) + \theta_0.
\end{equation}
Calculating the dissipation rate for this director gives
\begin{equation} \label{eqn:DefectEnergyDissipation}
    \int \gamma \left(\frac{d \theta}{dt}\right)^2\,dA = \gamma\pi m^2\log\left(\frac{R}{\varepsilon}\right)|u_0|^2
\end{equation}
where $R$ is a large scale cutoff (either the system size or some characteristic defect length such as the average distance between defects) and $\varepsilon$ the defect core size.
This yields the friction coefficient $\xi =  \gamma\pi m^2\log(R/\varepsilon)$.

Combining the drag force and elastic force, we have the velocity of the $j$th defect: in general
\begin{equation}
    u_j = -\frac{K}{\gamma}\frac{2}{m_j \log(R/\varepsilon)} i\left.\partial_{\zb}\theta_{\text{ext}}\right|_{z=z_j}.
\end{equation}
For a collection of defects with the same winding magnitude $|m|$, the velocity of the $j$th defect due to the rest of the defects is
\begin{equation} \label{eqn:DefectForceVelocityQuasiStatic}
    u_j = \frac{K}{\gamma\log(R/\varepsilon)}\sum_{k\neq j}\frac{\sgn(m_jm_k)}{\zb_j - \zb_k},
\end{equation}
yielding a $u \sim 1/d$ dependence on the distance between defects, which has been measured in experiments \cite{deGennes75,zushi22}.
For two defects of opposite winding, Eq.~\eqref{eqn:DefectForceVelocityQuasiStatic} predicts that their distance as a function of time scales as $d(t)\sim (t_0 - t)^{1/2}$ where $t_0$ is the annihilation time \cite{zushi22}.

The above calculation of the drag force assumed a quasi-static director approximation.
However, as was originally pointed out by Ryskin and Kremenetsky \cite{Ryskin91}, and followed up on by Denniston \cite{Denniston96}, Radzihovsky \cite{Radzihovsky15}, and more recently Romano, Mahault, and Golestanian \cite{Romano23,Romano24}, this assumption leads to the unphysical result that the friction coefficient diverges in the limit of an infinite system (due to the $\log(R)$ dependence).
This unphysical result may be relieved by substituting the moving defect director, Eq.~\eqref{eqn:MovingDefectDirector}, into the expression for the director dissipation rate.
Since the moving defect distortion is only felt by the texture inside a horizon of size $K/(\gamma |u_0|)$, the dissipation no longer diverges at large system sizes.
Importantly, the moving defect distortion renormalizes the friction coefficient so that 
\begin{equation} \label{eqn:AnomalousFriction}
    \xi(|u_0|) = \gamma\pi m^2\log\left(\frac{3.7 K}{\gamma \varepsilon |u_0|}\right).
\end{equation}
The renormalized friction coefficient no longer carries a divergence with the system size, but instead depends on the speed of the defect $|u_0|$, leading to an anomalous defect mobility (``anomalous'' in the sense that the velocity is not linearly related to the force).
Equation \eqref{eqn:AnomalousFriction} is often written in terms of a generalized Ericksen number, $\mathcal{E} = \gamma |u_0| \varepsilon/K$, which characterizes the ratio of the (rotational) viscous force to the elastic force on the moving defect.
The friction coefficient is then $\xi = \gamma\pi m^2 \log(3.7/\mathcal{E})$.
We note that the factor of $3.7$ appearing in the expression for the friction comes from the integration; it is an approximation of the symbolic expression  $4e^{1/2 - \gamma_E}$ where $\gamma_E$ is the Euler-Mascheroni constant \cite{Romano24}.
This factor only applies to the case of a defect moving at a constant velocity in one direction.
The numerical factor will differ depending on the actual motion of the defect, as explored in more detail in Refs.~\cite{Romano23,Romano24}.

Finally, we briefly mention the effect of spiral charges on this methodology.
In all of the above, it was assumed that the only singularity at the defect core was of the form $\arg(z)$. 
If singularities of the form $\log(|z|)$ (which carry zero winding number) are added as boundary conditions at the defect core, the computation of the force and the velocity-dependent director field encounter divergences in the integration.
Even if one assumes that the spiral charge terms are replaced by constant terms in the near field (as explored in Sec.~\ref{subsec:NearFieldDefectCore}), there is still a question of whether and how this should impact the elastic force on a defect.
That is, assuming Eq.~\eqref{eqn:DefectElasticForce} gives the elastic force on a defect, differentiating the energy (after integration) including spiral charges, Eq.~\eqref{eqn:SpiralChargeFreeEnergy}, leads to an elastic force that is central, yet it is known from numerical simulations that nonzero spiral charge induces a transverse motion in defects \cite{vromans16,tang17,schimming23}.
As explored in Refs.~\cite{Romano23,Romano24}, the spiral charges may be related to a measure of angular momentum of a defect: if a defect's displacement at some time $t$ relative to time $t'$ is not co-linear with the velocity at $t'$, then a logarithmic contribution to the director configuration (dependent on the history of the defect's trajectory) is added.
Thus, for a full picture of defect dynamics including spiral charges, one not only needs to handle changes in effective momentum (i.e. force balance) but also changes in effective angular momentum (i.e. torque balance).
To our knowledge, such a complete picture has not yet been established in the literature, though definitions of the torque on a defect have been proposed for specific cases or from macroscopic, symmetry based arguments \cite{tang17,Tang19,Copar24}.

\subsection{Asymptotic Matching} \label{subsec:AsymptoticMatching}

In the above section, the physics of the defect core was essentially ignored, with the assumption that forces and dissipation in the core region (where the scalar order parameter $S$ is varying) are small compared with that of the director.
However, the expressions for the core should match up to expressions for the director field at some intermediate distance from the core for a moving defect.
This is the motivation behind the method of matched asymptotic expansions, developed in depth for the motion of topological defects by Pismen and Rodriguez \cite{Pismen90,pismen99} and applied more recently to active nematics \cite{Pismen2013,shankar18,Shankar19,Montagna26}(see Sec.~\ref{subsec:Hydro}).
We briefly review the ideas and basic results of the method here.

We start by assuming a defect is moving at a constant velocity $u_0$ due to some constant external phase gradient.
In particular, we assume the order parameter near the defect of interest is given by Eq.~\eqref{eqn:OrderParameterDecomp} with $S$ and $\theta_d$ satisfying near-field and far-field conditions, while $\theta_{\text{ext}} = \Re[\bar{k} z]$ results in a phase gradient of strength $|k|$ in the direction perpendicular to the motion of the defect (i.e. $\theta_{\text{ext}}$ is a spin-wave).
In this way, the velocity of the defect is assumed to be proportional to the phase gradient via $u_0 = i \nu k$, where $\nu$ is a real number (this is inspired by the expression for the elastic force on the defect, Eq.~\eqref{eqn:ForceExternalField}).

The velocity will be obtained by matching the inner limit of the far-field director, with the outer limit of the near-field director.
We already discussed the far-field director, which is given by Eq.~\eqref{eqn:MovingDefectDirector} (plus the external phase).
Expanding the Green's function in Eq.~\eqref{eqn:MovingDefectDirector} and computing terms to (log)-linear order in $|z|$ gives
\begin{multline} \label{eqn:InnerLimitOuterDirector}
    \theta(|z|,\varphi) \approx m\varphi \\ 
    - \frac{m |z|}{2}\log\left(\frac{|u_0||z|}{4}e^{\gamma_E - 1}\right)\Im[\bar{u}_0e^{i\varphi}] + \Re[\bar{k} z] \\
    = m\varphi - \frac{m\Im[\bar{u}_0 z]}{2}\log\left(\frac{|u_0||z|}{4}e^{\gamma_E - 1 - 2\nu/m}\right)
\end{multline}
where we have introduced a time scale, $\tau = \gamma/C$, which, along with the length scale $\xi = \sqrt{K/C}$, allows us to non-dimensionalize the expressions.

For the near field director, we must solve the near field equations, Eqs.~\eqref{eqn:NearCoreS} and \eqref{eqn:NearCoreTheta}, in the rest frame of the moving defect:
\begin{align}
    2\Re[u_0\partial_z S] + 4\partial_z\partial_{\zb} S + S(1 - S^2 - 16|\partial_z\theta |^2) &= 0 \label{eqn:MovingNearCoreS} \\
    \Re[u_0\partial_z\theta] + 2\partial_z\partial_{\zb}\theta + \frac{8 }{S}\Re[\partial_z S\partial_{\zb}\theta] &= 0 \label{eqn:MovingNearCoreTheta}
\end{align}
Although the full equations are complicated, the important feature of this method is that only asymptotic solutions ($|z|\to \infty$) are required.
As such, we seek only asymptotic solutions to Eqs.~\eqref{eqn:MovingNearCoreS} and \eqref{eqn:MovingNearCoreTheta}.
An \textit{ansatz} for such an asymptotic solution is \cite{pismen99},
\begin{align}
    S(|z|,\varphi) &= S_0(|z|) + \alpha(|z|) \Re[\bar{u}_0e^{i\varphi}] \label{eqn:MovingSAsymptote} \\
    \theta(r,\varphi) &= m\varphi + \beta(|z|)\Im[\bar{u}_0e^{i\varphi}] \label{eqn:MovingThetaAsymptote}
\end{align}
where $S_0(|z|)$ satisfies the static near field equation, Eq.~\eqref{eqn:DefectCoreEquation}.
Substituting Eqs.~\eqref{eqn:MovingSAsymptote} and \eqref{eqn:MovingThetaAsymptote} in Eqs.~\eqref{eqn:MovingNearCoreS} and \eqref{eqn:MovingNearCoreTheta} and keeping terms to linear order in $|\mathbf{v}|$ yields
\begin{align}
    \alpha'' + \frac{1}{|z|}\alpha' + \left(1 - \frac{4m^2+1}{|z|^2}-3S_0^2\right)\alpha + \frac{8 m S_0}{|z|^2}\beta &= -S_0' \label{eqn:SAsymptoteExpansion} \\
    \beta'' + \frac{1}{|z|}\beta' - \frac{1}{|z|^2}\beta + \frac{4}{S_0}\left(S_0'\beta' + \frac{m\alpha}{|z|^2}\right) &= -\frac{m}{r} \label{eqn:ThetaAsymptoteExpansion}
\end{align}
where $'$ denotes a derivative with respect to $|z|$.

In the large $|z|$ limit, we have $S_0 \to 1$ and $\alpha \to 0$ so Eq.~\eqref{eqn:ThetaAsymptoteExpansion} becomes a simple linear inhomogeneous equation for $\beta$.
An asymptotic solution is $\beta(|z|) = -(1/2)m |z|\log(|z|/a)$ where $a$ is a yet undetermined constant.
The asymptotic director angle is thus given by
\begin{equation} \label{eqn:OuterLimitInnerDirector}
    \theta(|z|,\varphi) = m\varphi - \frac{m \Im[\bar{u}_0 z]}{2}\log\left(\frac{|z|}{a}\right).
\end{equation}
Requiring that Eqs.~\eqref{eqn:OuterLimitInnerDirector} and \eqref{eqn:InnerLimitOuterDirector} coincide sets
\begin{equation}
    \frac{1}{a} = \frac{|u_0|}{4 }e^{\gamma_E - 1 - 2\nu/m}.
\end{equation}
Solving the above expression for the phase gradient determines the mobility of the defect.
The relationship between the velocity and the phase gradient is then
\begin{equation} \label{eqn:VelocityGradientRelationship}
    \frac{m}{2}\log\left(\frac{1}{ \tilde{a} |u_0|}\right)u_0 = -ik
\end{equation}
where $\tilde{a} = (a/4)e^{\gamma_E-1}$.
The value of $\tilde{a}$ depends on the exact problem of interest and the structure of the core. 
For a defect moving at a slow, constant velocity, one may determine the value of $\tilde{a}$ directly by solving Eq.~\eqref{eqn:SAsymptoteExpansion}, or by satisfying solvability conditions and computing the energy of the core \cite{pismen99}.
Ref.~\cite{pismen99} computes $\tilde{a} \approx 0.304$ for the defect moving at a constant velocity.

We note that the form of Eq.~\eqref{eqn:VelocityGradientRelationship} is qualitatively similar to the velocity determined in the previous section using force balance and only the director.
The conceptual difference is that the core is now explicitly handled via the constant $\tilde{a}$ which may be either computed numerically, or approximated analytically (e.g. by assuming a core structure like the Pad\'e approximate in Eq.~\eqref{eqn:PadeApproximateCore}).
The result is similar since we assumed that forces are mediated by phase gradients, or spin waves (or, if the reader likes, ``Goldstone bosons'').
Therefore, for a defect in the presence of other defects, the velocity is predicted to be a solution of the transcendental equation:
\begin{equation}
    \log\left(\frac{1}{\tilde{a} |u_j|}\right)u_j = \sum_{k\neq j}\frac{\sgn(m_jm_k)}{\zb_j - \zb_k}.
\end{equation}
The velocity is qualitatively equivalent to Eq.~\eqref{eqn:DefectForceVelocityQuasiStatic} if the velocity dependent logarithmic mobility is replaced by a constant mobility (appropriate for defects sufficiently far apart in the quasi-static limit).

While the asymptotic matching approach is more technically and conceptually difficult than the force balance approach, it leads to a reasonable approximation of the defect velocities using regularized quantities (namely, the near-field order parameter), rather than the singular director field alone.
The ability to handle the shape of the core to various orders in $u_0$ has been shown to be helpful in complicated problems, such as in active nematics or elastically anisotropic nematics \cite{Pismen2013,shankar18,Montagna26}.
The question of how adding spiral charges affects the method and its results is still, to our knowledge, an open question.
Since the near core equations with spiral charges in the static case have not yet been sorted out, it is not obvious how to proceed when the defects are moving.

\subsection{Global Dynamics from Quasi-static Inertial Manifold} \label{subsec:GlobalDynamics}

Another method for approximating the defect velocity, particularly in the case when there are multiple defects in the system, also takes advantage of the product form of the order parameter, Eq.~\eqref{eqn:OrderParameterDecomp}.
This method, developed by Vafa \textit{et al.} \cite{Vafa20,Vafa22}, attempts to determine defect velocities (as well as orientations, or other intrinsic properties of defects) directly from the order parameter evolution equation, Eq.~\eqref{eqn:BerisEdwards}.
While the method was originally developed to understand defect dynamics in active nematics (see Sec.~\ref{subsec:Hydro}) we will summarize the methodology and results here assuming the fluid velocity $v = 0$.
The overall idea behind the method is to assume a particular form for the order parameter $q$ so that it is explicitly a function of defect positions and, potentially, orientations or other intrinsic defect properties.
The simplest form is the quasi-static configuration, which is taken in Refs.~\cite{Vafa20,Vafa22}.
The defect velocities then fall out from the time evolution of the order parameter.

We begin by writing the order parameter of a configuration in a particular form.
As mentioned above, if a system of many defects is dilute (i.e. defects are well separated) the total order parameter of the system may be written as a product of local defect order parameters.
In particular, near the $j$th defect in the system, one may write the quasi-static order parameter
\begin{equation} \label{eqn:QuasistaticOrderParameter}
    q_0(z,\zb;\{z_k(t)\}) = S(|z - z_j|)\left(\frac{z-z_j}{\zb-\zb_j}\right)^{m_j}\prod_{k\neq j}\left(\frac{z - z_k}{\zb - \zb_k}\right)^{m_k}
\end{equation}
where $S(|z - z_j|)$ satisfies Eq.~\eqref{eqn:NearCoreS} and the coordinates of the defects are the time varying quantities.
The approximation thus assumes that the time evolution of the order parameter proceeds only on the subspace parameterized by the positions of the defects.
Refs.~\cite{Vafa20,Vafa22} call this subspace the \textit{inertial manifold}.

If we now assume that, at any snapshot in time, $q(z,\zb,t) = q_0(z,\zb,\{z_k(t)\})$, we may rewrite the order parameter evolution equation in terms of the defect velocities since
\begin{equation}
    \partial_t q = \sum_k \left(v_k\partial_{z_k}q_0 + \bar{v}_k\partial_{\zb_k}q_0\right).
\end{equation}
Since we are assuming no fluid flow, the evolution equation for the order parameter Eq.~\eqref{eqn:BerisEdwards} reduces to $\partial_tq = -(1/\gamma)\delta \mathcal{F}/\delta \bar{q}$.
The variational derivative of the free energy in terms of $q_0$ is
\begin{equation}
    \frac{\delta \mathcal{F}}{\delta{\bar{q}_0}} = \sum_k\left(\frac{\delta z_k}{\delta \bar{q}_0}\frac{\delta \mathcal{F}}{\delta z_k} + \frac{\delta \zb_k}{\delta \bar{q}_0}\frac{\delta \mathcal{F}}{\delta \zb_k}\right).
\end{equation}
The quantity $\delta z_k /\delta \bar{q}_0$ is defined by ensuring that it is the inverse (in a variational sense) of $\delta \bar{q}_0 / \delta z_{\ell}$ \cite{Vafa20}.
Requiring this leads to
\begin{align}
    \frac{\delta z_k}{\delta \bar{q}_0} &= \sum_j\mathcal{M}_{kj}^{-1} \partial_{\zb_j}q_0 \\
    \mathcal{M}_{k\ell} &= \int_{\Gamma} \partial_{\zb_k}q_0\partial_{z_{\ell}}\bar{q}_0 \, dz d\zb . \label{eqn:MatrixMobilityDef}
\end{align}
Note that in the above treatment we have ignored the contribution of $\delta \zb_k/\delta \bar{q}_0$.
One may treat this term by extending the above argument; however, it is found that the resulting matrix (call it $\mathcal{N}_{k\ell}$) is negligible compared to $\mathcal{M}_{k \ell}$ in the limit of well separated defects and large system size (which is the limit in which the approximation represented by \eqref{eqn:QuasistaticOrderParameter} holds in the first place) \cite{Vafa20,Vafa22}.

Putting these together, and using the fact that $\mathcal{M}_{k\ell}$ is Hermitian, we arrive at
\begin{equation}
    \sum_{k}u_k \partial_{z_k}q_0 = -\frac{1}{\gamma}\sum_{\ell}\sum_{k}\mathcal{M}^{-1}_{k\ell}\partial_{\zb_\ell}\mathcal{F} \partial_{z_k}q_0.
\end{equation}
Comparing coefficients on $\partial_{z_k}q_0$ gives a velocity-force relationship
\begin{equation}
    u_k = \sum_\ell \frac{\mathcal{M}_{k\ell}^{-1}}{\gamma}F_{\text{elastic},\ell}
\end{equation}
where we have used Eq.~\eqref{eqn:DefectElasticForce} as the definition of the elastic force on a defect.
The result is an anisotropic mobility (or friction) coefficient that did not appear in the previous methodologies.
The non-centrality and non-locality of the velocity-force relationship (that is, that the velocity of defect $k$ depends on the forces on all the defects in the system) is due to the constraint that the order parameter always be described by the $q_0$ of Eq.~\eqref{eqn:QuasistaticOrderParameter}, the quasi-static inertial manifold.
Determining where and how fast a given defect will move requires knowledge of the forces on all the defects in order to remain within the quasistatic approximation.
Other interpretations may be found in Refs.~\cite{Vafa20,Vafa22}.

The friction coefficient matrix may be computed for a system of $N$ defects of size $R$ by substituting Eq.~\eqref{eqn:QuasistaticOrderParameter} into Eq.~\eqref{eqn:MatrixMobilityDef}.
The result is
\begin{equation}
    \mathcal{M}_{k\ell} = 4\pi m_k m_{\ell}\begin{cases}
        \log\left(\frac{R}{|z_k - z_\ell|}\right) & k\neq \ell \\
        \log\left(\frac{R}{a}\right) & k=\ell
    \end{cases}
\end{equation}
where $a\sim O(\varepsilon)$ is a factor that depends on the structure of $S(|z|)$ which may be computed numerically or approximated from, for example, the Pad\'e approximant in Eq.~\eqref{eqn:PadeApproximateCore}.
Similar to the force balance equations with a quasistatic approximation, the friction coefficients diverge with the system size, and indeed have a similar form.
Since the off-diagonal components of $\mathcal{M}$ scale as $\log(R/|z_k - z_\ell|)$, they are only logarithmically smaller than the diagonal components (since we assume $|z_k - z_\ell| \gg a$).
The velocity of the $j$th defect is then
\begin{equation}
    u_j = \frac{\pi K}{\gamma}\sum_{\ell}^N\mathcal{M}_{j\ell}^{-1}\sum_{k\neq \ell}^N\frac{m_k m_{\ell}}{\zb_\ell - \zb_k}.
\end{equation}

Finally, we mention defect orientations may, in principle, be additional parameters and incorporated in the approximate inertial manifold.
This is explored briefly in Refs.~\cite{Vafa20,Vafa22}, primarily in the limit of large activity.
The resulting dynamics for the full system would likely lead to coupled equations for the defect position and orientation, though we are unaware of any work that has pursued such a calculation in this framework.

\subsection{Halperin-Mazenko Topological Charge Conservation} \label{subsec:HalperinMazenko}

The previous section discussed a global method of determining defect velocities directly from the order parameter evolution equations.
Here we discuss instead a local method, in which one tracks individual defect positions as zeros of the order parameter.
Tracking zeros of the order parameter results in an expression for the conservation of topological charge, which then allows one to solve for defect velocities in terms of the evolution of the order parameter at the defect location.

The methodology as applied to nematics was first implemented by Angheluta \textit{et al.} \cite{angheluta21} to understand the motion of defects in active nematics.
However, the general method for defects in the Ginzburg-Landau model was first described by Halperin \cite{halperin81} and then later developed further by Mazenko in a series of papers \cite{liu92,mazenko97,mazenko99,Mazenko01}.
We review the method as it applies to nematics here.

The starting point is to realize that, if we are deep in the nematic phase, zeros of $q$ will correspond to defect locations.
The defect density, $\rho(z,\zb,t) = \sum_j m_j \delta[z - z_j(t)]$, can then be transformed into an expression in terms of the order parameter:
\begin{equation} \label{eqn:DefectDensity}
    \rho(z,\zb,t) = \sum_j m_j \delta[z - z_j(t)] = D(z,\zb,t)\delta[q(t)]
\end{equation}
where $D$ is the Jacobian determinant of the transformation from coordinates $z,\zb$ to $q,\bar{q}$.
$D$ may be written in terms of the order parameter,
\begin{equation}
    D(z,\zb,t) = \partial_zq\partial_{\zb}\bar{q} - \partial_z \bar{q} \partial_{\zb} q.
\end{equation}
It is important to note that $D$ carries both the sign and magnitude of the defect charges, that is, integrating $D$ over an area will give the total winding number in that area \cite{blow14,schimming22,Schimming24c}.
Another important feature of $D$ is that it obeys a conservation law in terms of a flux $J$:
\begin{align}
    \partial_t D &= 2\Re\left[\partial_z J\right] \\
    J &= \partial_t q \partial_{\zb}\bar{q} - \partial_t \bar{q}\partial_{\zb} q.
\end{align}

The defect density is also a conserved quantity, satisfying $\partial_t\rho = 2\Re\left[\partial_z J_{\rho}\right]$.
The flux $J_{\rho}$ may be computed in two different ways.
Differentiating the first equality in Eq.~\eqref{eqn:DefectDensity} with respect to time (and using the appropriate properties of the delta function \cite{liu92,mazenko97}) gives
\begin{equation}
    J_{\rho} = -\sum_j m_j u_j\delta\left[z - z_j\right].
\end{equation}
On the other hand, differentiating the second equality in Eq.~\eqref{eqn:DefectDensity} with respect to time yields \cite{liu92,mazenko97}
\begin{equation}
    J_{\rho} = J\delta\left[q\right] = \sum_j m_j \frac{J}{D}\delta\left[z - z_j\right]
\end{equation}
where the second equality inverts Eq.~\eqref{eqn:DefectDensity} to express $\delta\left[q\right]$ in terms of $\delta\left[z - z_j\right]$.
Comparing the expressions for defect density flux gives an expression for the defect velocity:
\begin{equation} \label{eqn:HalperinMazenkoVel}
    u_j = -\left.\frac{J}{D}\right|_{z=z_j} = \left.\frac{\partial_t \bar{q} \partial_{\zb} q - \partial_t q \partial_{\zb} \bar{q}}{\partial_{z} q \partial_{\zb} \bar{q} - \partial_z \bar{q}\partial_{\zb} q}\right|_{z=z_j}.
\end{equation}
The delta functions in the expressions for defect density flux require that $J/D$ be computed at the location of the defect core.

Equation \eqref{eqn:HalperinMazenkoVel} gives an expression for the defect velocity in terms of derivatives of the order parameter computed at the defect core.
It is exact in the sense that if $q(z,\zb,t)$ (along with its space and time derivatives) is known then Eq.~\eqref{eqn:HalperinMazenkoVel} will give the resulting velocity, which can be useful for determining defect velocities from order parameter data.
In order to give an analytical prediction of defect velocities, we must make some approximations of the local defect structure.
We first assume that near the $j$th defect, the order parameter is given by
\begin{equation} \label{eqn:qApproximationExternal}
    q(z,\zb) = S(|z-z_j|) \left(\frac{z -z_j}{\zb - \zb_j}\right)^{m_j} e^{2i\theta_{\text{ext}}},
\end{equation}
essentially Eq.~\eqref{eqn:OrderParameterDecomp} where we have assumed a quasi-static director with no spiral charge.
Importantly, since $J$ and $D$ are only required to be calculated at the defect core, we may take a local order parameter which satisfies Eqs.~\eqref{eqn:NearCoreS} and \eqref{eqn:NearCoreTheta} in the limit $|z - z_j| \to 0$.
We will thus assume that $S(|z|)$ is given by the linear core approximation, $S(|z|) = |z|/\varepsilon$ for $|z| < \varepsilon$. A more accurate approximation of the core structure such as the Pad\'e approximant Eq.~\eqref{eqn:PadeApproximateCore} will only change the results by a small numerical factor, that is, the velocity will not scale differently with the core size if a different approximation is used.

To find defect velocities at a given snapshot in time we must also assume a model for the time evolution of the order parameter.
As in the above section, we assume $q$ evolves according to the Beris-Edwards model, Eq.~\eqref{eqn:BerisEdwards}.
In the limit of zero fluid flow, this reduces to $\partial_t q = -(1/\gamma)\delta \mathcal{F}/\delta\bar{q}$, where we further assume that the free energy is given by the Landau-de Gennes free energy with a single elastic constant, Eq.~\eqref{eqn:Landau-deGennes}.
This leaves $\partial_t q = (K/\gamma) \nabla^2 q + f(q)$ where $f(q)$ is a function of $q$ but not derivatives of $q$.
At the defect core $f(q)$ goes to zero so we may ignore this term.
Substituting this into Eq.~\eqref{eqn:HalperinMazenkoVel} yields
\begin{equation} \label{eqn:HalperinMazenkoExternal}
    u_j = -\frac{16 K m_j}{\gamma}i\left.\partial_{\zb}\theta_{\text{ext}}\right|_{z=z_j}.
\end{equation}

For a system of $N$ defects under the quasi-static approximation the Halperin-Mazenko method predicts the velocity of the $j$th defect to be (we have explicitly assumed $m_k = \pm 1/2$)
\begin{equation} \label{eqn:QuasiStaticHalperinMazenko}
    u_j = 2\frac{K}{\gamma}\sum_{k\neq j}\frac{\sgn(m_j m_k)}{\zb_j - \zb_k}.
\end{equation}
Compare this to the velocity obtained from force balance, Eq.~\eqref{eqn:DefectForceVelocityQuasiStatic}.
The expressions differ only by a constant factor, though we note that the Halperin-Mazenko method leads to an expression that does not diverge with system size.
An advantage of the Halperin-Mazenko method (as well as the previous global inertial manifold method) is that one does not need to assume any forms of the force on a given defect, nor does one need to assume that the velocity is proportional to an external phase gradient, as in the asymptotic matching method.
Instead, the dynamics of defects fall out naturally from the expression for the time evolution of the order parameter.
A disadvantage is the need for an accurate expression for the order parameter at the core of each defect, which may not be readily available depending on the history of trajectories, presence of fluid flow, elastic constants, or boundary conditions. 

We note that since the Halperin-Mazenko method only requires calculation of the order parameter at the defect core, one may easily handle configurations with non-zero spiral charges.
As explored above, the near core limit of the spiral charge configuration simply adds a constant to the near field director $\theta_d$.
As such, the calculation remains unchanged and Eq.~\eqref{eqn:HalperinMazenkoExternal} still predicts the defect velocity (in the limit of zero flow, single elastic constant, and small defect speed).
For a system described by Eq.~\eqref{eqn:MultiDefectSpiralCharge}, the velocity is computed to be
\begin{equation}
    u_j = 8\frac{K}{\gamma}\sum_{k\neq j} \left[\frac{m_j m_k}{\zb_j - \zb_k} - i\frac{m_j \mu_k}{\zb_j - \zb_k}\right].
\end{equation}
Importantly, the factor of $i$ on the spiral charge term shifts the velocity by $\pi/2$ from the direction of the defect centers.
This transverse component of the velocity is observed in simulations of defects with nonzero spiral charge \cite{vromans16,tang17,pearce21,schimming23,Copar24}.

Finally, while we have provided examples for relatively simple cases where the director is described by its quasi-static description or perturbations thereof, we emphasize that the above methods may also be used when the local director field has a different form, such as for the case of periodic boundary conditions (i.e. Eq.~\eqref{eqn:PeriodicDirector}).
In these cases, to obtain accurate and physical predictions of the velocity, one must use the local director field associated with the correct free energy and boundary conditions.
This may substantially change the form of the predicted velocity! 
For one example, the director field associated with the Weierstrass $\sigma$ function for periodic boundary conditions predicts an asymmetric interaction between defects in a periodic domain (which can be attributed to topological solitons) \cite{Schimming25b}.
For another example, if the order parameter for the one elastic constant approximation is used, the above methods would predict that a $+1/2$ defect would self-propel when the system is elastically anisotropic ($K_{11}\neq K_{33}$) even for a passive liquid crystal \cite{schimming23,Romano24,Montagna26}.
However, as shown recently by Montagna \textit{et al.}, if the local order parameter is obtained using perturbation theory, the more physical result that $+1/2$ defects do not exert forces on themselves may be recovered \cite{Montagna26}. 

\subsection{Example Comparison between Methods}

\begin{figure}
\centering
    \includegraphics[width = \columnwidth]{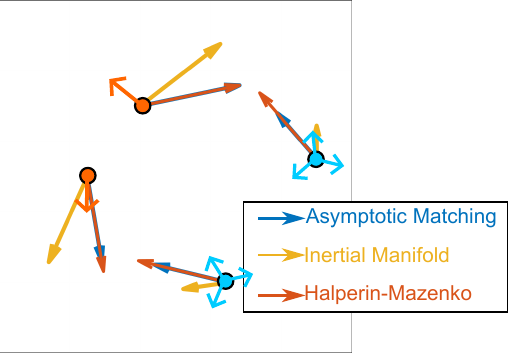}
    \caption{Defect velocities predicted by asymptotic matching (Sec.~\ref{subsec:AsymptoticMatching}), the global inertial manifold  method (Sec.~\ref{subsec:GlobalDynamics}), and the Halperin-Mazenko formalism (Sec.~\ref{subsec:HalperinMazenko}) for a system of two $+1/2$ defects at locations $z = -0.5L$ and $z = -0.2L + i0.4$ and two $-1/2$ defects at locations $z = 0.3L -i0.6L$ and $z = 0.8L + i0.1L$ where $L$ is the length of the field of view in the figure.
    For the inertial manifold method, we assumed a large-scale cutoff $R/L=10$ and small-scale cutoff $a/L = 0.01$.
    Defect positions and orientations are indicated with bright orange points and arrows for $+1/2$ defects and light blue points and threefold arrows for $-1/2$ defects.
    We assume spiral charges $\mu_i = 0$ so that defect orientations are determined by their positions.
    } 
    \label{fig:VelocityComparison}
\end{figure}

In Fig.~\ref{fig:VelocityComparison} we plot an example comparison between the defect velocities predicted by the methods of Secs.~\ref{subsec:AsymptoticMatching}, \ref{subsec:GlobalDynamics}, and \ref{subsec:HalperinMazenko} for interacting defects.
In the example we place two defects with winding $m=+1/2$ at $z = -0.5L$ and $z = -0.2L +i0.4L$ and two defects with winding $m=-1/2$ at $z = 0.3L -i0.6L$ and $z = 0.8L + 0.1L$ where $L$ is the length of the field of view of the figure.
We also assume a large-scale cutoff $R/L = 10$ and small-scale cutoff $a/L = 0.01$, which are needed to approximate velocities from the inertial manifold method.
We further assume that spiral charges are zero.

While the asymptotic matching and Halperin-Mazenko methods produce results that are qualitatively similar (that is, defect velocities in the same direction with different magnitudes), the inertial manifold method predicts defects of like sign move at a slightly steeper angle away from one another due to the non-centrality of the effective forces.
We speculate that one of the reasons for the difference in directions between the inertial manifold method and the other two is that defect orientations are not allowed to change as a result of motion in the inertial manifold method.
Thus, there may be a more limited set of directions that defects are able to move under the assumptions of the inertial manifold method.
We note that a more comprehensive comparison between methods which includes comparison to numerical and experimental data would be necessary to better understand the differences in assumptions and limitations of each method.

\subsection{Hydrodynamics and Active Nematics} \label{subsec:Hydro}

We conclude our look at defect dynamics by adding a flow field and asking how that affects the motion of the defects.
In all of the above sections we assumed the background flow of the nematic $v=0$.
However, even in an overdamped (low Reynolds number) setting, elastic stresses must balance viscous stresses, resulting in flow structures that drag and deflect defects.
Additionally, many of the recent advances in our understanding of defect dynamics have come about because of the recent surge in active nematic research.
In these inherently out-of-equilibrium systems, strong active stresses (generated by microscopic energy consumption) enter into the mix of force balance and lead to the self-propulsion of positive defects in nematics!
Here we review some of the recent analytical results regarding how underlying flow changes defect dynamics.

The flow field that appears in the Beris-Edwards equation Eq.~\eqref{eqn:BerisEdwards} may be thought of as some externally imposed flow to the liquid crystal.
However, due to the large amounts of elastic stress near defects, internal viscous stresses balancing these elastic stresses often create large flow structures themselves, termed ``backflow.''
As such, to fully understand the effect of hydrodynamics, one must solve the Navier-Stokes equation, or at least the Stokes equation for low Reynolds number flow.
In complex coordinates this is
\begin{equation} \label{eqn:StokesEq}
    4\eta\partial_z\partial_{\zb}v - \kappa v - 2\partial_{\zb}\Pi + 2\partial_{\zb}\Theta + 2\partial_{z}\Sigma = 0
\end{equation}
where $\eta$ is the fluid viscosity, $\kappa$ is a substrate friction coefficient, $\Pi$ is the hydrostatic pressure, and $\Theta$ and $\Sigma$ are complex-valued forms of the stress tensor $\bm{\sigma}$, explicitly written as
\begin{align}
    \Theta &= \frac{1}{2}\left[\sigma_{xx} + \sigma_{yy} - i(\sigma_{xy} - \sigma_{yx})\right] \\
    \Sigma &= \frac{1}{2}\left[\sigma_{xx} - \sigma_{yy} + i(\sigma_{xy} + \sigma_{yx})\right].
\end{align}
The stress in passive nematics is given by an elastic stress and the Ericksen stress, which can be more conveniently written in terms of the tensor order parameter $\mathbf{Q}$ and the variational derivative of the free energy $\mathbf{H} = -\delta\mathcal{F}/\delta\mathbf{Q}$:
\begin{align}
    \bm{\sigma}^{\text{elastic}} &= \lambda\mathbf{H} + \mathbf{Q}\mathbf{H} - \mathbf{H}\mathbf{Q} \\
    \bm{\sigma}^{\text{Ericksen}} &=-\nabla\mathbf{Q} : \frac{\partial f_e}{\partial(\nabla\mathbf{Q})} 
\end{align}

There has been limited analytical work on the effect of backflow on the motion of defects in nematics due to the complexity of the resulting equations (they are nonlinear in the order parameter and second order in derivatives).
Some recent approaches have used a variational method combined with perturbation theory to predict how various terms in the elastic stress contribute to the local flow field around defects and the resulting motion \cite{Sonnet05,Sonnet09,Tang19}.
The general, qualitative conclusions are that backflow increases the speed of $+1/2$ winding defects while it slows $-1/2$ defects, an effect that has also been observed in numerical studies \cite{toth02,Sonnet05}.

More recently, there has been significant interest in the effect of active stresses on the flow field near defects.
Active stresses occur due to energy input at the molecular scale, where the motion of microscopic constituents generates force dipoles.
Coarse-graining these force dipoles leads to a simple active stress written in terms of the tensor order parameter \cite{Simha02,Thampi16}
\begin{equation} \label{eqn:ActiveStress}
    \bm{\sigma}^{\text{active}} = -\zeta \Q
\end{equation}
where $\zeta$ is a constant that determines the direction and magnitude of the coarse-grained dipolar force, where $\zeta > 0$ refers to extensile active force dipoles (``pushers'') while $\zeta < 0$ refers to contractile active force dipoles (``pullers''). 
We note that Eq.~\eqref{eqn:ActiveStress} can also be established from symmetry arguments as the lowest order stress that has the symmetry of a local dipolar stress along the local orientation of the nematic.
Higher order terms and their effects on the flow around defects have been explored, though only to a limited degree analytically \cite{Maitra18,Zhao24,Zhao25}.

Substituting Eq.~\eqref{eqn:ActiveStress} into the Stokes equation, Eq.~\eqref{eqn:StokesEq}, and dropping elastic and Ericksen stresses gives an equation for the velocity field that is linear in both $v$ and $q$:
\begin{equation} \label{eqn:ActiveStokes}
    4\eta\partial_{z}\partial_{\zb}v - \kappa v - 2\partial_{\zb}\Pi -2\zeta\partial_z q = 0.
\end{equation}
We note also that the fluid is often assumed to be incompressible; the resulting constraint $2\Re[\partial_z v] = 0$ determines the pressure gradient in the above equation \cite{doo18}.
By combining Eq.~\eqref{eqn:ActiveStokes} with Eq.~\eqref{eqn:BerisEdwards} the various analytical techniques for describing defect motion may be used.

From Eq.~\eqref{eqn:ActiveStokes}, a length scale $\ell_h = \sqrt{\eta/\kappa}$ called the hydrodynamic screening length may be defined.
The screening length characterizes the range of hydrodynamic interactions in the system.
In particular, it is common to take one of two approximations: either $\ell_h \gg R$ (a ``wet'' active nematic) or $\ell_h \ll R$ (a ``dry'' nematic) where $R$ is the system size or some other macroscopic characteristic length (such as the average distance between defects).
Dry active nematics are easier to deal with analytically since one simply sets $\eta = 0$ in Eq.~\eqref{eqn:ActiveStokes}.
This leads to a flow velocity that is proportional to the divergence of the incompressible part of the active stress, allowing for direct substitution into the Beris-Edwards equation, Eq.~\eqref{eqn:BerisEdwards}, so that there is a single equation for the time evolution of $q$.
Recent studies have applied force balance \cite{Miyazako25}, asymptotic matching \cite{shankar18,Shankar19,Montagna26}, inertial manifold \cite{Vafa20,Vafa22}, and Halperin-Mazenko \cite{angheluta21} to dry active nematics.
The results are qualitatively similar: $+1/2$ defects are self-propelled along their symmetry axis (the direction depending on the sign of $\zeta$), while $-1/2$ defects are essentially passive, and only move as the result of passive elastic forces.
The benefit of studying a dry active nematic is that one never needs to compute the actual velocity field, although one sees from all of the calculations that defects are simply advected by the local fluid velocity.

\begin{figure}
\centering
    \includegraphics[width = \columnwidth]{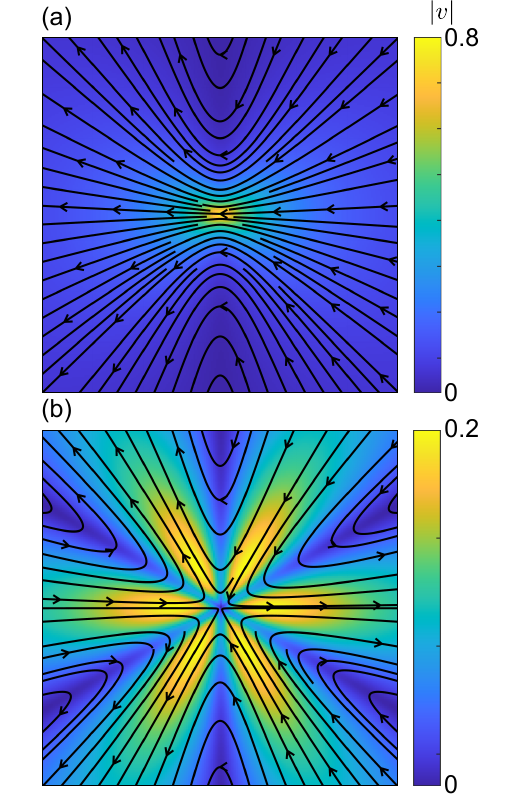}
    \caption{Solutions for the flow velocity, Eq.~\eqref{eqn:ActiveStokes} around a (a) $+1/2$ winding defect and (b) $-1/2$ winding defect.
    The lines show streamlines of the flow solution while the color shows the magnitude.
    Note that for the $+1/2$ defect the magnitude is maximized at the core while for the $-1/2$ defect the magnitude is zero at the core.
    The field of view for both configurations goes from $-10\sqrt{\eta/\kappa}$ to $10\sqrt{\eta/\kappa}$ in both the $x$ and $y$ axes, and set the activity $\zeta = 1$ so that it is extensile.
    } 
    \label{fig:FlowVelocities}
\end{figure}

For a wet active nematic, one must solve the Stokes equation (with $\kappa=0$) for the fluid velocity near a defect core before substituting the fluid velocity into Eq.~\eqref{eqn:BerisEdwards}.
This was first computed in Ref.~\cite{giomi14} using a Green's function.
We note that solutions to the Stokes equation diverge in two dimensions, so a large-scale cut-off must be used for the integrals to converge.
The finding was that positive $+1/2$ defects support two counter-rotating vortices along their symmetry axis with nonzero fluid velocity at the core of the defect, while negative defects support six vortices distributed around the defect, but the fluid velocity goes to zero at the core due to the sixfold symmetry (see Fig.~\ref{fig:FlowVelocities}).
Assuming that the defect velocity is equal to the fluid velocity at the defect core yields the same qualitative picture as studies of dry active nematics: $+1/2$ defects are self-propelled while $-1/2$ defects are not. 
In addition to directly solving Eq.~\eqref{eqn:ActiveStokes}, Ref.~\cite{Pismen2013} uses the method of asymptotic matching (with a calculation very similar to that of Sec.~\ref{subsec:AsymptoticMatching}) applied to the wet limit of the active nematic.
The findings are qualitatively identical, that $+1/2$ defects are self-propelled, while $-1/2$ defects are not, though we note that no large-scale cut-off is needed in the asymptotic matching approach.

While it is common to either use the dry limit or wet limit, recent calculations have shown that it is not necessary to choose one or the other.
R\o{}nning \textit{et al.} computed solutions to Eq.~\eqref{eqn:ActiveStokes} around $+1/2$ and $-1/2$ defect cores for nonzero $\eta$ and $\kappa$ \cite{ronning22}.
The results are qualitatively similar to those of Ref.~\cite{giomi14}, but do not require a large-scale cut-off since the hydrodynamic screening length, $\ell_h = \sqrt{\eta/\kappa}$, provides a length scale for the decay of the hydrodynamic solutions.
We show the streamlines and magnitude of the flow in Fig.~\ref{fig:FlowVelocities}.
As is predicted by other methods, the flow is finite (actually maximized) at the core of $+1/2$ winding defects, while it is zero at the core of $-1/2$ winding defects.
In the figure, we show the solutions out to $10\ell_h$ in each coordinate direction and set $\zeta = 1$ (so that the dynamics are extensile).
The resulting flow at the core of the $+1/2$ defect results in a propulsion direction toward its ``comet head'' (compare with the defect oriented as in Fig.~\ref{fig:DefectTopology}).

The flow solutions in Fig.~\ref{fig:FlowVelocities} were recently used with the Halperin-Mazenko formalism to derive equations for defect motion in active nematics in infinite and confined geometries by approximating the global fluid flow as a linear superposition of individual defect flows \cite{Schimming25,Matsukiyo26}.
This method, though highly approximate, was able to reproduce defect behaviors seen in computational studies and experiments, such as $+1/2$ defect co-rotation \cite{norton18,opathalage19,sokolov24}, and defect braiding \cite{Memarian24,Klein26}.

There is much more that could be said about active nematics and the role defects play in them.
As mentioned earlier, our goal here was to briefly mention the primary recent analytical contributions to our understanding of defect motion.
There are now several reviews on active nematics which cover defect motion \cite{doo18,Brezin22,Shankar22,Yeomans23} and which give a broader look at the field of active nematics as a whole.

\section{Future Challenges and Conclusion} \label{sec:Conclusion}

We have examined multiple analytical results and methods regarding the structure of two-dimensional nematic defects, and shown how one may go from equations for the structure to equations for the defect velocity.
There are still many questions that remain from an analytical perspective that may further our understanding and allow us to more precisely program both synthetic and biological applications involving nematic defects.

While we have many tools to analytically describe the structure of defected configurations in nematics (most of which were reviewed in Sec.~\ref{sec:DefectStructure}), there are still cases where analytical descriptions elude us.
One of the most important cases are systems with anisotropic elasticity, that is $K_{11}\neq K_{33}$ in the Oseen-Frank elastic energy.
Because of the nonlinearity of Eq.~\eqref{eqn:AnisoELEquation} describing configurations, general solutions for multiple defects are unknown.
This has consequences when, for example, attempting to infer elastic constants from local defect geometries \cite{hudson89}.
It also has consequences for predicting the dynamics of these defects, since accurate near field directors are required to obtain accurate defect velocity predictions \cite{Romano23,schimming23,Romano24,Montagna26}.

Another class of important two-dimensional scenarios with few fully analytical results are curved surfaces.
Closed surfaces such as the sphere or torus have constraints on the total winding due to the Poincar\'e-Hopf theorem, though director configurations with arbitrary defects are, to our knowledge, not known analytically except in simple cases such as the ground state on a sphere (which must have a total winding of $2$) \cite{Vitelli06}.
While there are several numerical studies generating textures of both polar and nematic fields on curved surfaces \cite{Selinger2011-gm,Dhakal12,ellis18,Nitschke18,Pearce19,Pearce20,Napoli21,Rojo-Gonzalez24}, analytical descriptions of the director are scant, especially in cases with arbitrarily positioned topological defects.
These descriptions are important in the ever-evolving field of active nematics, since many of the experimental systems are biological systems which evolve on explicitly curved surfaces.
Both extrinsic and intrinsic curvature are known to affect defect motion and stability \cite{Selinger2011-gm,ellis18,Pearce19,Rojo-Gonzalez24}, thus having analytical descriptions of the director would allow greater understanding of the mechanisms to manipulate defect interactions.

Another aspect of defect motion that has, in our view, not received enough attention is the coupling of defect motion and defect orientation.
Often, as in Refs.~\cite{Vafa20,Vafa22,angheluta21,Schimming25}, the orientation is assumed to be given by the quasi-static director field, and so it is determined by the positions of defects.
In other studies, such as in Refs.~\cite{Pearce19,shankar18,Shankar19}, the orientation is an independent variable, but is evolved in time with a random noise, not coupled to the actual motion of the defect (that is, defects are treated as active Brownian particles).
In principle, the local defect orientation should be independently tuned by adding variable spiral charge to each defect.
How the spiral charge evolves, however, should be in some way coupled to the translational motion of the defects.
This is what is observed in numerical studies: defects with initial spiral charge (giving them non-optimal orientation) move transverse to one another until the orientation becomes optimal and the spiral charge of each defect is zero; they then continue either annihilating or repelling along the separation line of the two defects \cite{vromans16,tang17,schimming23}.
The closest analytical calculation of how the spiral charge evolves in time is in Ref.~\cite{Romano24}, where spiral charges are related to trajectory dependent measures of the effective angular momentum of each defect.
The resulting expression is nonlocal-in-time, and thus difficult to use in an effective particle theory of defects.
To our knowledge, no local-in-time approximation of the change in spiral charge has been put forward.
That is, we are not aware of any local equations of motion for the $\mu_i$ in Eq.~\eqref{eqn:MultiDefectSpiralCharge}.
Such descriptions would be invaluable to the goal of deriving a ``particle model'' of defect dynamics.

In the previous section we briefly discussed the role of hydrodynamics and how backflow and active stresses may influence defect motion.
As we mentioned, there have been many calculations describing the resulting fluid flow around topological defects due to the active stress, but far fewer (and no full solutions, to our knowledge) describing the elastic backflow near a defect.
Analytical calculations of the backflow would be invaluable for our understanding of defect interactions and microfluidic control of defect configurations.
They would also allow us to better understand the importance of various terms in the elastic stress, as well as better measure various material parameters of the fluid.

Finally, we have limited the discussion here to two-dimensional nematics; defects in three dimensions promise to provide even richer defect configurations and dynamics, with applications to both passive and active systems, and probably deserve a review of their own.
There are also far fewer analytical studies of defect structure and dynamics in three dimensions owing to the highly nonlinear equations governing the systems.
Indeed, even problems as simple as describing the minimal energy defect director for two arbitrary defects in three dimensions is unsolved, to our knowledge.
Approximate solutions for the director structure around defects as well as defect dynamics (using force balance methods \cite{long21,Long24}, the Halperin-Mazenko formalism \cite{schimming22,schimming23,Zushi24}, or topological methods \cite{Pollard25,Pollard25b}) have been explored, yet quantitative, and even qualitative, verification of the predictions remain scarce.
Additionally, several studies have focused on flow solutions near defects applied to three-dimensional active nematics \cite{binysh20,Houston22,Houston23}, though how these flows affect defect interactions is still being addressed.

We hope that this review will aid in the furthering of analytical approaches to nematic defect structure and motion and that it may inspire new and creative methods for describing and predicting defect behaviors.

\begin{acknowledgements}
This material is based upon work supported in part by the National Science Foundation under Grant No. DMR-2225543. CDS also gratefully acknowledges support from the Department of Physics and Astronomy at Johns Hopkins University through the William H. Miller III postdoctoral fellowship.

\end{acknowledgements}


\bibliography{LC}

@book{deGennes75,
    author = "de Gennes, P. G.",
	title = "The Physics of Liquid Crystals",
	year = "1975",
	publisher = "Oxford University Press"
}

@misc{Klein2026Signatures,
author = {Klein, Brandon and Steinbock, Chanania and Murali, Aniruddh and Amey, Christopher and Baskaran, Aparna and Mitchell, Kevin A. and Beller, Daniel A.},
title = {Signatures of elastic anisotropy in non-equilibrium nematic textures},
howpublished = {In preparation},
year = {2026}
}

@book{beris94,
    author = "Beris, Antony N. and Edwards, Brian J.",
    title  = "Thermodynamics of flowing systems",
    year = "1994",
    publisher = "Oxford University Press"
}

@article{alexander12,
    author = "Alexander, Gareth P. and Chen, Bryan Gin-ge and Matsumoto, Elisabetta and Kamien, Randall D.",
    title = "Colloquium: Disclination loops, point defects, and all that in nematic liquid crystals",
    journal = "Rev. Mod. Phys.",
    volume = "84",
    pages = "497",
    year = "2012",
    doi = "10.1103/RevModPhys.84.497"
    
}

@article{zhou17,
    author = "Zhou, Shuang and Shiyanovskii, Sergij V. and Park, Heung-Shik and Lavrentovich, Oleg D.",
	title = "Fine structure of the topological defect cores studied for disclinations in lyotropic chromonic liquid crystals",
	journal = "Nature Commun.",
	volume = "8",
	pages = "14974",
	year = "2017"
}

@book{pismen99,
	author = "Pismen, L. M.",
	title = "Vortices in Nonlinear Fields",
	year = "1999",
	publisher = "Oxford University Press"
}

@article{frank58,
	author = "Frank, F.C.",
	title = "On the theory of liquid crystals",
	journal = "Discuss. Faraday Soc.",
	volume = "25",
	pages = "19",
	year = "1958"
}

@article{longa87,
	author = "Longa, L. and Monselesan, D. and Trebin, H. -R.",
	title = "An extension of the {L}andau-{G}inzberg-de {G}ennes theory for liquid crystals",
	journal = "Liq. Cryst.",
	volume = "2",
	pages = "769",
	year = "1987"
}

@article{sen86,
	author = "Sen, A. K. and Sullivan, D. E.",
	title = "{L}andau-de {G}ennes theory of wetting and orientational transitions at a nematic-liquid--substrate interface",
	journal = "Phys. Rev. A",
	volume = "35",
	pages = "1391",
	year = "1987"
}

@article{selinger19,
	author = "Jonathan V. Selinger",
	title = "Interpretation of saddle-splay and the {O}seen-{F}rank free energy in liquid crystals",
	journal = "Liquid Crystals Reviews",
	volume = "6",
	number = "2",
	pages = "129-142",
	year  = "2018",
    doi = "10.1080/21680396.2019.1581103"
}

@article{gramsbergen86,
	title = "{L}andau theory of the nematic-isotropic phase transition",
	journal = "Phys. Rep.",
	volume = "135",
	number = "4",
	pages = "195 - 257",
	year = "1986",
	issn = "0370-1573",
	author = "Egbert F. Gramsbergen and Lech Longa and Wim H. de Jeu"
}

@article{lyuksyutov78,
	title = "Topological instability of singularities at small distances in nematics",
	journal = "Sov. Phys. JETP",
	author = "Lyuksyutov, I. F.",
	volume = "48",
	pages = "178",
	year = "1978"
}

@article{angheluta21,
    title = "The role of fluid flow in the dynamics of active nematic defects",
    author = "Angheluta, Luiza and Chen, Zhitao and Marchetti, M. Cristina and Bowick, Mark J.",
    journal = "New J. Phys.",
    volume = "23",
    pages = "033009",
    year = "2021",
    doi = "10.1088/1367-2630/abe8a8"
}

@article{long21,
    author = "Long, Cheng and Tang, Xingzhou and Selinger, Robin L.B. and Selinger, Jonathan V.",
    title = "Geometry and mechanics of disclination lines in 3D nematic liquid crystals",
    journal = "Soft Matter",
    volume = "17",
    pages = "2265",
    year = "2021",
    doi = "10.1039/D0SM01899F"
}

@article{schopohl87,
	title = "Defect core structure in nematic liquid crystals",
	author = "Schopohl, N. and Sluckin, T.J.",
	journal = "Phys. Rev. Lett.",
	volume = "59",
	pages = "22",
	year = "1987",
    doi = "10.1103/PhysRevLett.59.2582"
}

@article{dzyaloshinskii70,
	title = "Theory of disinclinations in liquid crystals",
	author = "Dzyaloshinskii, I. E.",
	journal = "Sov. Phys. JETP",
	volume = "31",
	pages = "773-780",
	year = "1970"
}

@misc{mottram14,
	author = "Mottram, Nigel J. and Newton, Christopher J.P.",
	title = "Introduction to {Q}-tensor theory",
	howpublished = "e-print arXiv:1409.3542v2 [cond-mat.soft]",
	year = "2014"
}

@article{binysh20,
	author="Binysh, Jack and Kos, {\u{Z}}iga and \u{C}opar, Simon and Ravnik, Miha and Alexander, Gareth P.",
	title = "Three-Dimensional Active Defect Loops",
	journal="Phys. Rev. Lett.",
	year="2020",
	volume="124",
	pages="088001"
}

@article{Houston22,
  title = {Defect loops in three-dimensional active nematics as active multipoles},
  author = {Houston, Alexander J. H. and Alexander, Gareth P.},
  journal = {Phys. Rev. E},
  volume = {105},
  issue = {6},
  pages = {L062601},
  numpages = {6},
  year = {2022},
  month = {Jun},
  publisher = {American Physical Society},
  doi = {10.1103/PhysRevE.105.L062601},
  url = {https://link.aps.org/doi/10.1103/PhysRevE.105.L062601}
}

@article{schimming20b,
  title={Anisotropic disclination cores in nematic liquid crystals modeled by a self-consistent molecular field theory},
  author={Schimming, Cody D and Vi{\~n}als, Jorge},
  journal={Phys. Rev. E},
  volume={102},
  number={1},
  pages={010701(R)},
  year={2020},
  publisher={APS},
  doi = {10.1103/PhysRevE.102.010701},
}

@article{hudson89,
	author = "Hudson, Steven D. and Thomas, Edwin L.",
	title = "Frank Elastic-Constant Anisotropy Measured from Transmission-Electron-Microscope Images of Disclinations",
	journal = "Phys. Rev. Lett.",
	volume = "62",
	pages = "1993",
	year = "1989"
}

@article{shankar18,
	author = "Shankar, Suraj and Ramaswamy, Sriram and Marchetti, M. Cristina and Bowick, Mark J.",
	title = "Defect Unbinding in Active Nematics",
	journal = "Phys. Rev. Lett.",
	volume = "121",
	pages = "108002",
	year = "2018"
}

@article{doo18,
	author = "Doostmohammadi, A. and Ign\'{e}s-Mullol, J. and Yeomans, J. M. and Sagu\'{e}s, F.",
	title = "Active Nematics",
	journal = "Nature Commun.",
	volume = "9",
	pages = "3246",
	year = "2018"
}

@article{paparini21,
title="Nematic tactoid population",
author="Paparini, Silvia and Virga, Epifanio G.",
journal="Phys. Rev. E",
volume="103",
pages="022707",
year="2021"
}

@incollection{halperin81,
    author="Halperin, B. I.",
    editor = "Balian, Roger",
    booktitle = "Physics of Defects",
    publisher = "North-Holland Pub. Co.",
    year = "1981"
}

@article{liu92,
    author = "Liu, Fong and Mazenko, Gene F.",
    title = "Defect-defect correlation in the dynamics of first-order phase transitions",
    journal="Phys. Rev. B",
    volume = "46",
    pages = "5963",
    year = "1992"
}

@article{mazenko97,
    author = "Mazenko, Gene F. and Wickham, Robert A.",
    title = "Ordering kinetics of defect structures",
    journal = "Phys. Rev. E",
    volume = "57",
    pages = "2539",
    year = "1997"
}

@article{zushi22,
author = {Yohei Zushi  and Kazumasa A. Takeuchi },
title = {Scaling and spontaneous symmetry restoring of topological defect dynamics in liquid crystal},
journal = {Proc. Natl. Acad. Sciences},
volume = {119},
number = {41},
pages = {e2207349119},
year = {2022},
doi = {10.1073/pnas.2207349119},
}

@article{vromans16,
    author = "Vromans, Arthur J. and Giomi, Luca",
    title = "Orientational properties of nematic disclinations",
    journal = "Soft Matter",
    volume = "12",
    pages = "6490",
    year = "2016"
}

@article{mazenko99,
    author = "Mazenko, Gene F.",
    title = "Velocity distribution for strings in phase-ordering kinetics",
    journal = "Phys. Rev. E",
    volume = "59",
    pages = "1574",
    year = "1999"
}

@article{tang17,
    author = "Tang, Xingzhou and Selinger, Jonathan V.",
    title = "Orientation of topological defects in 2D nematic liquid crystals",
    journal = "Soft Matter",
    volume = "13",
    pages = "5481",
    year = "2017",
    doi = "10.1039/C7SM01195D"
}

@article{blow14,
    author = "Blow, Matthew L. and Thampi, Sumesh P. and Yeomans, Julia M.",
    title = "Biphasic, Lyotropic, Active Nematics",
    journal = "Phys. Rev. Lett.",
    volume = "113",
    pages = "248303",
    year = "2014"
}

@article{stark01,
    author = "Stark, Holger",
    title = "Physics of colloidal dispersions in nematic liquid crystals",
    journal = "Phys. Rep.",
    volume = "351",
    pages = "387-474",
    year = "2001",
    doi = "10.1016/S0370-1573(00)00144-7"
}

@article{pearce21,
author = "Pearce, D. J. G. and Kruse, K.",
title = "Properties of twisted topological defects in 2D nematic liquid crystals",
journal = "Soft Matter",
volume = "17",
pages = "7408",
year = "2021"
}

@article{bowick94,
author = {Mark J. Bowick  and L. Chandar  and E. A. Schiff  and Ajit M. Srivastava },
title = {The Cosmological Kibble Mechanism in the Laboratory: String Formation in Liquid Crystals},
journal = {Science},
volume = {263},
number = {5149},
pages = {943-945},
year = {1994}
}

@article{toth02,
  title = {Hydrodynamics of Topological Defects in Nematic Liquid Crystals},
  author = {T\'oth, G\'eza and Denniston, Colin and Yeomans, J. M.},
  journal = {Phys. Rev. Lett.},
  volume = {88},
  issue = {10},
  pages = {105504},
  year = {2002},
}

@article{ellis18,
    author = "Ellis, Perry W. and Pearce, Daniel J. G. and Chang, Ya-Wen and Goldsztein, Guillermo and Giomi, Luca and Fernandez-Nieves, Alberto",
    title = "Curvature-induced defect unbinding and dynamics in active nematic toroids",
    journal = "Nature Physics",
    volume = "14",
    pages = "85--90",
    year = "2018"
}

@article{paparini21b,
    author = "Paparini, Silvia and Virga, Epifanio G.",
    title = "Shape bistability in 2D chromonic droplets",
    journal = "J. Phys. Condens. Matter",
    volume = "33",
    pages = "495101",
    year = "2021"
}

@article{kleman08,
  title = {Disclinations, dislocations, and continuous defects: A reappraisal},
  author = {Kleman, M. and Friedel, J.},
  journal = {Rev. Mod. Phys.},
  volume = {80},
  issue = {1},
  pages = {61--115},
  numpages = {0},
  year = {2008},
  month = {Jan},
  publisher = {American Physical Society},
  doi = {10.1103/RevModPhys.80.61},
  url = {https://link.aps.org/doi/10.1103/RevModPhys.80.61}
}

@article{opathalage19,
author = {Opathalage, A. and Norton, M. M. and Juniper, M. P. N. and Langeslay, B. and Aghvami, S. A. and Fraden, S. and Dogic, Z.},
title = {Self-organized dynamics and the transition to turbulence of confined active nematics},
journal = {Proc. Natl. Acad. Sci. (USA)},
volume = {116},
number = {11},
pages = {4788-4797},
year = {2019},
doi = {10.1073/pnas.1816733116},
}

@Article{schimming22,
author ="Schimming, Cody D. and Viñals, Jorge",
title  ="Singularity identification for the characterization of topology{,} geometry{,} and motion of nematic disclination lines",
journal  ="Soft Matter",
year  ="2022",
volume  ="18",
issue  ="11",
pages  ="2234-2244",
publisher  ="The Royal Society of Chemistry",
doi  ="10.1039/D1SM01584B",
}

@article{chuang91,
author = {Isaac Chuang  and Ruth Durrer  and Neil Turok  and Bernard Yurke },
title = {Cosmology in the Laboratory: Defect Dynamics in Liquid Crystals},
journal = {Science},
volume = {251},
number = {4999},
pages = {1336-1342},
year = {1991},
doi = {10.1126/science.251.4999.1336},
}

@article{kleman89,
doi = {10.1088/0034-4885/52/5/002},
url = {https://dx.doi.org/10.1088/0034-4885/52/5/002},
year = {1989},
month = {may},
publisher = {},
volume = {52},
number = {5},
pages = {555},
author = {M Kleman},
title = {Defects in liquid crystals},
journal = {Rep. Prog. Phys.},
}

@article{norton18,
  title = {Insensitivity of active nematic liquid crystal dynamics to topological constraints},
  author = {Norton, M. M. and Baskaran, A. and Opathalage, A. and Langeslay, B. and Fraden, S. and Baskaran, A. and Hagan, M. F.},
  journal = {Phys. Rev. E},
  volume = {97},
  issue = {1},
  pages = {012702},
  numpages = {8},
  year = {2018},
  month = {Jan},
  publisher = {American Physical Society},
  doi = {10.1103/PhysRevE.97.012702},
}

@article{giomi14,
author = {Giomi, L.  and Bowick, M. J. and Mishra, P. and Sknepnek, R. and Marchetti, M. C.},
title = {Defect dynamics in active nematics},
journal = {Phil. Trans. R. Soc. A},
volume = {372},
number = {2029},
pages = {20130365},
year = {2014},
doi = {10.1098/rsta.2013.0365},
URL = {https://royalsocietypublishing.org/doi/abs/10.1098/rsta.2013.0365},
}

@article{schimming23,
author = {Schimming, Cody D.  and Viñals, Jorge },
title = {Kinematics and dynamics of disclination lines in three-dimensional nematics},
journal = {Proc. R. Soc. A: Mathematical, Physical and Engineering Sciences},
volume = {479},
number = {2273},
pages = {20230042},
year = {2023},
doi = {10.1098/rspa.2023.0042},
}

@ARTICLE{Shankar19,
author = {Shankar, S. and Marchetti, M. C.},
title = {Hydrodynamics of Active Defects: From Order to Chaos to Defect Ordering},
journal = {Phys. Rev. X},
volume = {9},
issue = {4},
pages = {041047},
numpages = {18},
year = {2019},
doi = {10.1103/PhysRevX.9.041047},
}

@article{ronning22,
author = {R{\o}nning, J. and Marchetti, C. M. and Bowick, M. J. and Angheluta, L.},
title = {Flow around topological defects in active nematic films},
journal = {Proc. Roy. Soc. A: Math. Phys. Eng. Sci.},
volume = {478},
number = {2257},
pages = {20210879},
year = {2022},
doi = {10.1098/rspa.2021.0879},
}

@article{Maroudas-Sacks21,
    author = {Maroudas-Sacks, Yonit and Garion, Liora and Shani-Zerbib, Lital and Livshits, Anton and Braun, Erez and Keren, Kinneret},
    title = {Topological defects in the nematic order of actin fibres as organization centers of \textit{Hydra} morphogenesis},
    journal = {Nature Physics},
    year = {2021},
    volume = {17},
    pages = {251--259},
    doi = {10.1038/s41567-020-01083-1}
}

@article{Memarian24,
  title = {Controlling Chaos: Periodic Defect Braiding in Active Nematics Confined to a Cardioid},
  author = {Memarian, Fereshteh L. and Hammar, Derek and Sabbir, Md Mainul Hasan and Elias, Mark and Mitchell, Kevin A. and Hirst, Linda S.},
  journal = {Phys. Rev. Lett.},
  volume = {132},
  issue = {22},
  pages = {228301},
  numpages = {6},
  year = {2024},
  month = {May},
  publisher = {American Physical Society},
  doi = {10.1103/PhysRevLett.132.228301},
  url = {https://link.aps.org/doi/10.1103/PhysRevLett.132.228301}
}

@misc{sokolov24,
      title={Synthetic active liquid crystals powered by acoustic waves}, 
      author={Andrey Sokolov and Jaideep Katuri and Juan J. de Pablo and Alexey Snezhko},
      year={2024},
      eprint={2403.17268},
      archivePrefix={arXiv},
      primaryClass={cond-mat.soft},
      url={https://arxiv.org/abs/2403.17268}, 
}

@Article{Vafa22,
author ="Vafa, Farzan",
title  ="Defect dynamics in active polar fluids vs. active nematics",
journal  ="Soft Matter",
year  ="2022",
volume  ="18",
issue  ="42",
pages  ="8087-8097",
publisher  ="The Royal Society of Chemistry",
doi  ="10.1039/D2SM00830K",
url  ="http://dx.doi.org/10.1039/D2SM00830K",
}

@article{Pismen2013,
  title = {Dynamics of defects in an active nematic layer},
  author = {Pismen, L. M.},
  journal = {Phys. Rev. E},
  volume = {88},
  issue = {5},
  pages = {050502(R)},
  numpages = {5},
  year = {2013},
  month = {Nov},
  publisher = {American Physical Society},
  doi = {10.1103/PhysRevE.88.050502},
  url = {https://link.aps.org/doi/10.1103/PhysRevE.88.050502}
}

@article{Zushi24,
  title = {Approach and rotation of reconnecting topological defect lines in liquid crystal},
  author = {Zushi, Yohei and Schimming, Cody D. and Takeuchi, Kazumasa A.},
  journal = {Phys. Rev. Res.},
  volume = {6},
  issue = {2},
  pages = {023284},
  numpages = {10},
  year = {2024},
  month = {Jun},
  publisher = {American Physical Society},
  doi = {10.1103/PhysRevResearch.6.023284},
  url = {https://link.aps.org/doi/10.1103/PhysRevResearch.6.023284}
}

@Article{Schimming25,
author ="Schimming, Cody D. and Reichhardt, C. J. O. and Reichhardt, C.",
title  ="Analytical model for the motion and interaction of two-dimensional active nematic defects",
journal  ="Soft Matter",
volume = "21",
year  ="2025",
pages  ="122-136",
publisher  ="The Royal Society of Chemistry",
doi  ="10.1039/D4SM00956H",
url  ="http://dx.doi.org/10.1039/D4SM00956H",
}

@article{Mermin79,
  title = {The topological theory of defects in ordered media},
  author = {Mermin, N. D.},
  journal = {Rev. Mod. Phys.},
  volume = {51},
  issue = {3},
  pages = {591--648},
  numpages = {0},
  year = {1979},
  month = {Jul},
  publisher = {American Physical Society},
  doi = {10.1103/RevModPhys.51.591},
  url = {https://link.aps.org/doi/10.1103/RevModPhys.51.591}
}

@Article{Hill21,
	title={{Chiral magnetism: a geometric perspective}},
	author={Daniel Hill and Valeriy Slastikov and Oleg Tchernyshyov},
	journal={SciPost Phys.},
	volume={10},
	pages={078},
	year={2021},
	publisher={SciPost},
	doi={10.21468/SciPostPhys.10.3.078},
	url={https://scipost.org/10.21468/SciPostPhys.10.3.078},
}

@book{Whittaker_Watson_1996, 
place={Cambridge}, 
edition={4}, 
series={Cambridge Mathematical Library}, 
title={A Course of Modern Analysis}, 
publisher={Cambridge University Press}, 
author={Whittaker, E. T. and Watson, G. N.}, 
year={1996}, 
collection={Cambridge Mathematical Library}
}

@ARTICLE{Selinger2011-gm,
  title     = "Monte Carlo Studies of the {XY} Model on {Two-Dimensional}
               Curved Surfaces",
  author    = "Selinger, Robin L Blumberg and Konya, Andrew and Travesset, Alex
               and Selinger, Jonathan V",
  journal   = "J. Phys. Chem. B",
  publisher = "American Chemical Society",
  volume    =  115,
  number    =  48,
  pages     = "13989--13993",
  month     =  dec,
  year      =  2011
}

@article{Rojo-Gonzalez24,
  title = {Defect-populated configurations in nematic solid tori and cylinders},
  author = {Rojo-Gonz\'alez, Javier and Carenza, Livio Nicola and de la Cotte, Alexis and Hoffmann, Ludwig A. and Giomi, Luca and Fernandez-Nieves, Alberto},
  journal = {Phys. Rev. Res.},
  volume = {6},
  issue = {1},
  pages = {L012065},
  numpages = {6},
  year = {2024},
  month = {Mar},
  publisher = {American Physical Society},
  doi = {10.1103/PhysRevResearch.6.L012065},
  url = {https://link.aps.org/doi/10.1103/PhysRevResearch.6.L012065}
}

@Article{Copar24,
author ="Čopar, Simon and Kos, \v{Z}iga",
title  ="Many-defect solutions in planar nematics: interactions{,} spiral textures and boundary conditions",
journal  ="Soft Matter",
year  ="2024",
volume  ="20",
issue  ="35",
pages  ="6894-6906",
publisher  ="The Royal Society of Chemistry",
doi  ="10.1039/D4SM00586D",
}

@article{Chandler23, 
title={A nematic liquid crystal with an immersed body: equilibrium, stress and paradox}, 
volume={967}, 
DOI={10.1017/jfm.2023.488}, 
journal={Journal of Fluid Mechanics}, 
author={Chandler, Thomas G.J. and Spagnolie, Saverio E.}, 
year={2023}, 
pages={A19}}

@article{Chandler24,
author = {Chandler, Thomas G. J. and Spagnolie, Saverio E.},
title = {Exact and Approximate Solutions for Elastic Interactions in a Nematic Liquid Crystal},
journal = {SIAM Journal on Applied Mathematics},
volume = {84},
number = {6},
pages = {2476-2501},
year = {2024},
doi = {10.1137/23M1622908}
}

@article{Ginzburg58,
author = {Ginzburg, V. L. and Pitaevskii, L. P.},
title = {On the theory of superfluidity},
journal = {J. Exptl. Theoret. Phys.},
volume = {34},
pages = {1240--1245},
year = {1958},
}

@Article{Myers24,
author ="Myers, Lucas and Swift, Carter and Rønning, Jonas and Angheluta, Luiza and Viñals, Jorge",
title  ="A computational study of nematic core structure and disclination interactions in elastically anisotropic nematics",
journal  ="Soft Matter",
year  ="2024",
volume  ="20",
issue  ="13",
pages  ="2900-2914",
publisher  ="The Royal Society of Chemistry",
doi  ="10.1039/D3SM01616A",
url  ="http://dx.doi.org/10.1039/D3SM01616A",
}

@misc{Houston25,
      title={The Role of Elastic Anisotropy in Active Nematics}, 
      author={Alexander J. H. Houston},
      year={2025},
      eprint={2511.08520},
      archivePrefix={arXiv},
      primaryClass={cond-mat.soft},
      url={https://arxiv.org/abs/2511.08520}, 
}

@article{Lewis17,
author = {Lewis, A. H. and Aarts, D. G. A. L. and Howell, P. D. and Majumdar, A.},
title = {Nematic Equilibria on a Two-Dimensional Annulus},
journal = {Studies in Applied Mathematics},
volume = {138},
number = {4},
pages = {438-466},
doi = {https://doi.org/10.1111/sapm.12161},
year = {2017}
}

@Article{Schimming25b,
author ="Schimming, Cody D.",
title  ="Defect interactions through periodic boundaries in two-dimensional p-atics",
journal  ="Soft Matter",
year  ="2025",
volume  ="21",
issue  ="39",
pages  ="7673-7684",
publisher  ="The Royal Society of Chemistry",
doi  ="10.1039/D5SM00773A",
}

@misc{NIST:DLMF,
         key = "{\relax DLMF}",
       title = "{\it NIST Digital Library of Mathematical Functions}",
howpublished = "\url{https://dlmf.nist.gov/}, Release 1.2.6 of 2026-03-15",
         url = "https://dlmf.nist.gov/",
        note = "F.~W.~J. Olver, A.~B. {Olde Daalhuis}, D.~W. Lozier, B.~I. Schneider,
                R.~F. Boisvert, C.~W. Clark, B.~R. Miller, B.~V. Saunders,
                H.~S. Cohl, and M.~A. McClain, eds."}

@article{Kleman06,
author = {M. Kleman and O. D. Lavrentovich},
title = {Topological point defects in nematic liquid crystals},
journal = {Philosophical Magazine},
volume = {86},
number = {25-26},
pages = {4117--4137},
year = {2006},
publisher = {Taylor \& Francis},
doi = {10.1080/14786430600593016},
URL = {https://doi.org/10.1080/14786430600593016},
}

@article{Wu22,
author = {Jin-Sheng Wu and Ivan I. Smalyukh},
title = {Hopfions, heliknotons, skyrmions, torons and both abelian and nonabelian vortices in chiral liquid crystals},
journal = {Liquid Crystals Reviews},
volume = {10},
number = {1-2},
pages = {34--68},
year = {2022},
publisher = {Taylor \& Francis},
doi = {10.1080/21680396.2022.2040058},
URL = {https://doi.org/10.1080/21680396.2022.2040058},
}

@article{Nobili92,
  title = {Disorientation-induced disordering at a nematic-liquid-crystal--solid interface},
  author = {Nobili, M. and Durand, G.},
  journal = {Phys. Rev. A},
  volume = {46},
  issue = {10},
  pages = {R6174(R)--R6177(R)},
  numpages = {0},
  year = {1992},
  month = {Nov},
  publisher = {American Physical Society},
  doi = {10.1103/PhysRevA.46.R6174},
  url = {https://link.aps.org/doi/10.1103/PhysRevA.46.R6174}
}

@article{Fournier05,
doi = {10.1209/epl/i2005-10253-5},
url = {https://doi.org/10.1209/epl/i2005-10253-5},
year = {2005},
month = {oct},
publisher = {},
volume = {72},
number = {3},
pages = {403},
author = {J.-B. Fournier and P. Galatola},
title = {Modeling planar degenerate wetting and anchoring  in nematic
liquid crystals},
journal = {Europhysics Letters},
}

@article{Meyer69,
  title = {Piezoelectric Effects in Liquid Crystals},
  author = {Meyer, Robert B.},
  journal = {Phys. Rev. Lett.},
  volume = {22},
  issue = {18},
  pages = {918--921},
  numpages = {0},
  year = {1969},
  month = {May},
  publisher = {American Physical Society},
  doi = {10.1103/PhysRevLett.22.918},
  url = {https://link.aps.org/doi/10.1103/PhysRevLett.22.918}
}

@ARTICLE{James08,
  author={James, Richard and Willman, Eero and Anibal Fernandez, F. and Day, Sally E.},
  journal={IEEE Transactions on Magnetics}, 
  title={Computer Modeling of Liquid Crystal Hydrodynamics}, 
  year={2008},
  volume={44},
  number={6},
  pages={814-817},
  doi={10.1109/TMAG.2007.916029}}

@ARTICLE{Willman07,
  author={Willman, Eero and Fernandez, F. AnÍbal and James, Richard and Day, Sally E.},
  journal={IEEE Transactions on Electron Devices}, 
  title={Modeling of Weak Anisotropic Anchoring of Nematic Liquid Crystals in the Landau–de Gennes Theory}, 
  year={2007},
  volume={54},
  number={10},
  pages={2630-2637},
  doi={10.1109/TED.2007.904369}}

@article{Vitelli06,
  title = {Nematic textures in spherical shells},
  author = {Vitelli, V and Nelson, D. R.},
  journal = {Phys. Rev. E},
  volume = {74},
  issue = {2},
  pages = {021711},
  numpages = {18},
  year = {2006},
  month = {Aug},
  publisher = {American Physical Society},
  doi = {10.1103/PhysRevE.74.021711},
  url = {https://link.aps.org/doi/10.1103/PhysRevE.74.021711}
}

@Article{Tang19,
author ="Tang, Xingzhou and Selinger, Jonathan V.",
title  ="Theory of defect motion in 2D passive and active nematic liquid crystals",
journal  ="Soft Matter",
year  ="2019",
volume  ="15",
issue  ="4",
pages  ="587-601",
publisher  ="The Royal Society of Chemistry",
doi  ="10.1039/C8SM01901K",
}

@article{Dafermos70,
    author = {DAFERMOS, CONSTANTINE M.},
    title = {DISINCLINATIONS IN LIQUID CRYSTALS},
    journal = {The Quarterly Journal of Mechanics and Applied Mathematics},
    volume = {23},
    number = {2},
    pages = {49-64},
    year = {1970},
    month = {05},
    issn = {0033-5614},
    doi = {10.1093/qjmam/23.2.49},
    url = {https://doi.org/10.1093/qjmam/23.2.49},
    eprint = {https://academic.oup.com/qjmam/article-pdf/23/2/49/5240996/23-2-49.pdf},
}

@article{Imura73,
title = {Friction coefficient for a moving disinclination in a nematic liquid crystal},
journal = {Physics Letters A},
volume = {42},
number = {6},
pages = {403-404},
year = {1973},
issn = {0375-9601},
doi = {https://doi.org/10.1016/0375-9601(73)90728-7},
url = {https://www.sciencedirect.com/science/article/pii/0375960173907287},
author = {H. Imura and K. Okano},
}

@article{Ryskin91,
  title = {Drag force on a line defect moving through an otherwise undisturbed field: Disclination line in a nematic liquid crystal},
  author = {Ryskin, G. and Kremenetsky, M.},
  journal = {Phys. Rev. Lett.},
  volume = {67},
  issue = {12},
  pages = {1574--1577},
  numpages = {0},
  year = {1991},
  month = {Sep},
  publisher = {American Physical Society},
  doi = {10.1103/PhysRevLett.67.1574},
  url = {https://link.aps.org/doi/10.1103/PhysRevLett.67.1574}
}

@article{Denniston96,
  title = {Disclination dynamics in nematic liquid crystals},
  author = {Denniston, Colin},
  journal = {Phys. Rev. B},
  volume = {54},
  issue = {9},
  pages = {6272--6275},
  numpages = {0},
  year = {1996},
  month = {Sep},
  publisher = {American Physical Society},
  doi = {10.1103/PhysRevB.54.6272},
  url = {https://link.aps.org/doi/10.1103/PhysRevB.54.6272}
}

@article{Radzihovsky15,
  title = {Anomalous Energetics and Dynamics of Moving Vortices},
  author = {Radzihovsky, Leo},
  journal = {Phys. Rev. Lett.},
  volume = {115},
  issue = {24},
  pages = {247801},
  numpages = {4},
  year = {2015},
  month = {Dec},
  publisher = {American Physical Society},
  doi = {10.1103/PhysRevLett.115.247801},
  url = {https://link.aps.org/doi/10.1103/PhysRevLett.115.247801}
}

@article{Romano23,
doi = {10.1088/1742-5468/aceb57},
url = {https://doi.org/10.1088/1742-5468/aceb57},
year = {2023},
month = {aug},
publisher = {IOP Publishing},
volume = {2023},
number = {8},
pages = {083211},
author = {Romano, Jacopo and Mahault, Benoît and Golestanian, Ramin},
title = {Dynamical theory of topological defects I: the multivalued solution of the diffusion equation},
journal = {Journal of Statistical Mechanics: Theory and Experiment},
}

@article{Romano24,
doi = {10.1088/1742-5468/ad2ddb},
url = {https://doi.org/10.1088/1742-5468/ad2ddb},
year = {2024},
month = {mar},
publisher = {IOP Publishing},
volume = {2024},
number = {3},
pages = {033208},
author = {Romano, Jacopo and Mahault, Benoît and Golestanian, Ramin},
title = {Dynamical theory of topological defects II: universal aspects of defect motion},
journal = {Journal of Statistical Mechanics: Theory and Experiment},
}

@article{Pismen90,
  title = {Mobility of singularities in the dissipative Ginzburg-Landau equation},
  author = {Pismen, L. M. and Rodriguez, J. D.},
  journal = {Phys. Rev. A},
  volume = {42},
  issue = {4},
  pages = {2471(R)--2474(R)},
  numpages = {0},
  year = {1990},
  month = {Aug},
  publisher = {American Physical Society},
  doi = {10.1103/PhysRevA.42.2471},
  url = {https://link.aps.org/doi/10.1103/PhysRevA.42.2471}
}

@article{Montagna26,
  title = {Shape dictates the motion of topological defects in active nematics},
  author = {La Montagna, Giacomo Marco and Burekovi\ifmmode \acute{c}\else \'{c}\fi{}, Sumeja and Maitra, Ananyo and Nardini, Cesare},
  journal = {Phys. Rev. E},
  volume = {113},
  issue = {5},
  pages = {L053401},
  numpages = {7},
  year = {2026},
  month = {May},
  publisher = {American Physical Society},
  doi = {10.1103/b9g7-h378},
  url = {https://link.aps.org/doi/10.1103/b9g7-h378}
}

@misc{Vafa20,
      title={Multi-defect Dynamics in Active Nematics}, 
      author={Farzan Vafa and Mark J. Bowick and M. Cristina Marchetti and Boris I. Shraiman},
      year={2020},
      eprint={2007.02947},
      archivePrefix={arXiv},
      primaryClass={cond-mat.soft},
      url={https://arxiv.org/abs/2007.02947}, 
}

@article{Mazenko01,
  title = {Defect statistics in the two-dimensional complex Ginzburg-Landau model},
  author = {Mazenko, Gene F.},
  journal = {Phys. Rev. E},
  volume = {64},
  issue = {1},
  pages = {016110},
  numpages = {11},
  year = {2001},
  month = {Jun},
  publisher = {American Physical Society},
  doi = {10.1103/PhysRevE.64.016110},
  url = {https://link.aps.org/doi/10.1103/PhysRevE.64.016110}
}

@article{Balasubramaniam22,
title = {Active nematics across scales from cytoskeleton organization to tissue morphogenesis},
journal = {Current Opinion in Genetics \& Development},
volume = {73},
pages = {101897},
year = {2022},
issn = {0959-437X},
doi = {https://doi.org/10.1016/j.gde.2021.101897},
url = {https://www.sciencedirect.com/science/article/pii/S0959437X21001428},
author = {Lakshmi Balasubramaniam and René-Marc Mège and Benoît Ladoux},
}

@article{Wang06,
  title = {Bend-Induced Melting of the Smectic-$A$ Phase: Analogy to a Type-I Superconductor},
  author = {Wang, Ruiting and Syed, Ishtiaque M. and Carbone, Giovanni and Petschek, Rolfe G. and Rosenblatt, Charles},
  journal = {Phys. Rev. Lett.},
  volume = {97},
  issue = {16},
  pages = {167802},
  numpages = {4},
  year = {2006},
  month = {Oct},
  publisher = {American Physical Society},
  doi = {10.1103/PhysRevLett.97.167802},
  url = {https://link.aps.org/doi/10.1103/PhysRevLett.97.167802}
}

@Article{Crowdy06,
author={Crowdy, Darren
and Marshall, Jonathan},
title={Conformal Mappings between Canonical Multiply Connected Domains},
journal={Computational Methods and Function Theory},
year={2006},
month={Jun},
day={01},
volume={6},
number={1},
pages={59-76},
issn={2195-3724},
doi={10.1007/BF03321118},
url={https://doi.org/10.1007/BF03321118}
}

@article{Schimming24c,
    author = {Schimming, Cody D. and Vinals, Jorge},
    title = {A tensor density measure of topological charge in three-dimensional nematic phases},
    journal = {Proceedings of the Royal Society A: Mathematical, Physical and Engineering Sciences},
    volume = {480},
    number = {2290},
    pages = {20230564},
    year = {2024},
    month = {05},
    issn = {1364-5021},
    doi = {10.1098/rspa.2023.0564},
    url = {https://doi.org/10.1098/rspa.2023.0564},
    eprint = {https://royalsocietypublishing.org/rspa/article-pdf/doi/10.1098/rspa.2023.0564/512679/rspa.2023.0564.pdf},
}

@article{Sonnet09,
author = {André M. Sonnet and Epifanio G. Virga},
title = {Flow and reorientation in the dynamics of nematic defects},
journal = {Liquid Crystals},
volume = {36},
number = {10-11},
pages = {1185--1192},
year = {2009},
publisher = {Taylor \& Francis},
doi = {10.1080/02678290903034480},
}

@article{Napoli21,
    author = {Napoli, Gaetano and Pylypovskyi, Oleksandr V. and Sheka, Denis D. and Vergori, Luigi},
    title = {Nematic shells: new insights in topology- and curvature-induced effects},
    journal = {Soft Matter},
    volume = {17},
    number = {45},
    pages = {10322-10333},
    year = {2021},
    month = {12},
    issn = {1744-683X},
    doi = {10.1039/d1sm00719j},
}

@article{Pearce20,
doi = {10.1088/1367-2630/ab91fd},
url = {https://doi.org/10.1088/1367-2630/ab91fd},
year = {2020},
month = {jun},
publisher = {IOP Publishing},
volume = {22},
number = {6},
pages = {063051},
author = {Pearce, D J G},
title = {Defect order in active nematics on a curved surface},
journal = {New Journal of Physics},
}

@article{Nitschke18,
    author = {Nitschke, Ingo and Nestler, Michael and Praetorius, Simon and Löwen, Hartmut and Voigt, Axel},
    title = {Nematic liquid crystals on curved surfaces: a thin film limit},
    journal = {Proceedings of the Royal Society A: Mathematical, Physical and Engineering Sciences},
    volume = {474},
    number = {2214},
    pages = {20170686},
    year = {2018},
    month = {06},
    issn = {1364-5021},
    doi = {10.1098/rspa.2017.0686},
    url = {https://doi.org/10.1098/rspa.2017.0686},
    eprint = {https://royalsocietypublishing.org/rspa/article-pdf/doi/10.1098/rspa.2017.0686/357526/rspa.2017.0686.pdf},
}

@article{Dhakal12,
  title = {Nematic liquid crystals on spherical surfaces: Control of defect configurations by temperature, density, and rod shape},
  author = {Dhakal, Subas and Solis, Francisco J. and Olvera de la Cruz, Monica},
  journal = {Phys. Rev. E},
  volume = {86},
  issue = {1},
  pages = {011709},
  numpages = {9},
  year = {2012},
  month = {Jul},
  publisher = {American Physical Society},
  doi = {10.1103/PhysRevE.86.011709},
  url = {https://link.aps.org/doi/10.1103/PhysRevE.86.011709}
}

@article{Pearce19,
  title = {Geometrical Control of Active Turbulence in Curved Topographies},
  author = {Pearce, D. J. G. and Ellis, Perry W. and Fernandez-Nieves, Alberto and Giomi, L.},
  journal = {Phys. Rev. Lett.},
  volume = {122},
  issue = {16},
  pages = {168002},
  numpages = {5},
  year = {2019},
  month = {Apr},
  publisher = {American Physical Society},
  doi = {10.1103/PhysRevLett.122.168002},
  url = {https://link.aps.org/doi/10.1103/PhysRevLett.122.168002}
}

@article{Long24,
  title = {Frank-Read Mechanism in Nematic Liquid Crystals},
  author = {Long, Cheng and Deutsch, Matthew J. and Angelo, Joseph and Culbreath, Christopher and Yokoyama, Hiroshi and Selinger, Jonathan V. and Selinger, Robin L. B.},
  journal = {Phys. Rev. X},
  volume = {14},
  issue = {1},
  pages = {011044},
  numpages = {12},
  year = {2024},
  month = {Mar},
  publisher = {American Physical Society},
  doi = {10.1103/PhysRevX.14.011044},
  url = {https://link.aps.org/doi/10.1103/PhysRevX.14.011044}
}

@article{Pollard25b,
    author = {Pollard, Joseph and Morris, Richard G.},
    title = {Defect dynamics in cholesterics: beyond the Peach–Koehler force},
    journal = {Soft Matter},
    volume = {21},
    number = {12},
    pages = {2300-2316},
    year = {2025},
    month = {03},
    issn = {1744-683X},
    doi = {10.1039/d4sm01478b},
}

@article{Pollard25,
  title = {Morse Theory and Meron-Mediated Interactions Between Disclination Lines in Nematic Materials},
  author = {Pollard, Joseph and Morris, Richard G.},
  journal = {Phys. Rev. X},
  volume = {15},
  issue = {2},
  pages = {021099},
  numpages = {37},
  year = {2025},
  month = {Jun},
  publisher = {American Physical Society},
  doi = {10.1103/lbbj-txnx},
  url = {https://link.aps.org/doi/10.1103/lbbj-txnx}
}

@Article{Sonnet05,
author={Sonnet, Andr{\'e} M.},
title={Viscous forces on nematic defects},
journal={Continuum Mechanics and Thermodynamics},
year={2005},
month={Jul},
day={01},
volume={17},
number={3},
pages={287-295},
issn={1432-0959},
doi={10.1007/s00161-005-0204-x},
url={https://doi.org/10.1007/s00161-005-0204-x}
}

@article{Simha02,
  title = {Hydrodynamic Fluctuations and Instabilities in Ordered Suspensions of Self-Propelled Particles},
  author = {Aditi Simha, R. and Ramaswamy, Sriram},
  journal = {Phys. Rev. Lett.},
  volume = {89},
  issue = {5},
  pages = {058101},
  numpages = {4},
  year = {2002},
  month = {Jul},
  publisher = {American Physical Society},
  doi = {10.1103/PhysRevLett.89.058101},
  url = {https://link.aps.org/doi/10.1103/PhysRevLett.89.058101}
}

@Article{Thampi16,
author={Thampi, S. P.
and Yeomans, J. M.},
title={Active turbulence in active nematics},
journal={The European Physical Journal Special Topics},
year={2016},
month={Jul},
day={01},
volume={225},
number={4},
pages={651-662},
issn={1951-6401},
doi={10.1140/epjst/e2015-50324-3},
url={https://doi.org/10.1140/epjst/e2015-50324-3}
}

@article{Maitra18,
author = {Ananyo Maitra  and Pragya Srivastava  and M. Cristina Marchetti  and Juho S. Lintuvuori  and Sriram Ramaswamy  and Martin Lenz },
title = {A nonequilibrium force can stabilize 2D active nematics},
journal = {Proceedings of the National Academy of Sciences},
volume = {115},
number = {27},
pages = {6934-6939},
year = {2018},
doi = {10.1073/pnas.1720607115},
URL = {https://www.pnas.org/doi/abs/10.1073/pnas.1720607115},
eprint = {https://www.pnas.org/doi/pdf/10.1073/pnas.1720607115},
}

@article{Zhao24,
  title = {Integer Topological Defects Reveal Antisymmetric Forces in Active Nematics},
  author = {Zhao, Zihui and Yao, Yisong and Li, He and Zhao, Yongfeng and Wang, Yujia and Zhang, Hepeng and Chat\'e, Hugues and Sano, Masaki},
  journal = {Phys. Rev. Lett.},
  volume = {133},
  issue = {26},
  pages = {268301},
  numpages = {8},
  year = {2024},
  month = {Dec},
  publisher = {American Physical Society},
  doi = {10.1103/PhysRevLett.133.268301},
  url = {https://link.aps.org/doi/10.1103/PhysRevLett.133.268301}
}

@Article{Zhao25,
author={Zhao, Zihui
and Li, He
and Yao, Yisong
and Zhao, Yongfeng
and Serra, Francesca
and Kawaguchi, Kyogo
and Zhang, Hepeng
and Sano, Masaki},
title={Integer topological defects offer a methodology to quantify and classify active cell monolayers},
journal={Nature Communications},
year={2025},
month={Mar},
day={12},
volume={16},
number={1},
pages={2452},
doi={10.1038/s41467-025-57783-w},
}

@article{Miyazako25,
    author = {Miyazako, Hiroki and Miyoshi, Hiroyuki and Nara, Takaaki},
    title = {Defect dynamics in dry active nematics by residue calculus for holomorphic functions of nematic director field},
    journal = {Soft Matter},
    volume = {21},
    number = {29},
    pages = {5947-5956},
    year = {2025},
    month = {08},
    issn = {1744-683X},
    doi = {10.1039/d5sm00201j},
    url = {https://doi.org/10.1039/d5sm00201j},
    eprint = {https://pubs.rsc.org/sm/article-pdf/21/29/5947/10688826/d5sm00201j.pdf},
}

@article{Matsukiyo26,
  title = {Defect configuration of an active nematic around a circular obstacle},
  author = {Matsukiyo, Hiroki and Fukuda, Jun-ichi},
  journal = {Phys. Rev. E},
  volume = {113},
  issue = {2},
  pages = {025409},
  numpages = {13},
  year = {2026},
  month = {Feb},
  publisher = {American Physical Society},
  doi = {10.1103/zcbr-36t3},
  url = {https://link.aps.org/doi/10.1103/zcbr-36t3}
}

@article{Klein26,
author = {Brandon Klein  and Alejandro J. Soto Franco  and Md Mainul Hasan Sabbir  and Matthew J. Deutsch  and Ross Kliegman  and Robin L. B. Selinger  and Kevin A. Mitchell  and Daniel A. Beller },
title = {Chaos-generating periodic orbits of topological defects in confined active nematics},
journal = {Proceedings of the National Academy of Sciences},
volume = {123},
number = {28},
pages = {e2516670123},
year = {2026},
doi = {10.1073/pnas.2516670123},
URL = {https://www.pnas.org/doi/abs/10.1073/pnas.2516670123},
eprint = {https://www.pnas.org/doi/pdf/10.1073/pnas.2516670123},
}

@Article{Shankar22,
author={Shankar, Suraj
and Souslov, Anton
and Bowick, Mark J.
and Marchetti, M. Cristina
and Vitelli, Vincenzo},
title={Topological active matter},
journal={Nature Reviews Physics},
year={2022},
month={Jun},
day={01},
volume={4},
number={6},
pages={380-398},
issn={2522-5820},
doi={10.1038/s42254-022-00445-3},
url={https://doi.org/10.1038/s42254-022-00445-3}
}

@incollection{Yeomans23,
    author = {Yeomans, Julia M.},
    editor = {Kurzthaler, Christina and Gentile, Luigi and Stone, Howard A},
    isbn = {978-1-83916-229-9},
    title = {Active Nematics: Mesoscale Turbulence and Self-propelled Topological Defects},
    booktitle = {Out-of-equilibrium Soft Matter},
    publisher = {The Royal Society of Chemistry},
    year = {2023},
    month = {03},
    url = {https://doi.org/10.1039/9781839169465-00088},
}

@Article{Brezin22,
author={Br{\'e}zin, Louis
and Risler, Thomas
and Joanny, Jean-Francois},
title={Spontaneous flow created by active topological defects},
journal={The European Physical Journal E},
year={2022},
month={Apr},
day={07},
volume={45},
number={4},
pages={30},
issn={1292-895X},
doi={10.1140/epje/s10189-022-00186-2},
url={https://doi.org/10.1140/epje/s10189-022-00186-2}
}

@ARTICLE{Houston23,
    
AUTHOR={Houston, Alexander J. H.  and Alexander, Gareth P. },
           
TITLE={Active nematic multipoles: Flow responses and the dynamics of defects and colloids},
          
JOURNAL={Frontiers in Physics},
          
VOLUME={Volume 11 - 2023},
  
YEAR={2023},
  
URL={https://www.frontiersin.org/journals/physics/articles/10.3389/fphy.2023.1110244},
  
DOI={10.3389/fphy.2023.1110244},
  
ISSN={2296-424X},
}

@book{Arfken05,
  title={Mathematical Methods for Physicists},
  author={Arfken, George B. and Weber, Hans J.},
  edition={6th},
  year={2005},
  publisher={Elsevier Academic Press},
  address={Amsterdam; Boston},
  isbn={978-0-12-088584-8}
}

@book{Crowdy20,
author = {Crowdy, Darren},
title = {Solving Problems in Multiply Connected Domains},
publisher = {Society for Industrial and Applied Mathematics},
year = {2020},
doi = {10.1137/1.9781611976151},
address = {Philadelphia, PA},
edition   = {},
URL = {https://epubs.siam.org/doi/abs/10.1137/1.9781611976151},
eprint = {https://epubs.siam.org/doi/pdf/10.1137/1.9781611976151}
}

\end{document}